\documentclass[useAMS,usenatbib]{mn2e}
\usepackage{epsf}
\usepackage{epsfig}
\usepackage{subfigure}
\usepackage{amssymb}

\newcommand{\refsim}{\textsc{ref}}
\newcommand{\agn}{\textsc{agn}}
\newcommand{\nocool}{\textsc{nocool}}

\newcommand{\planck}{\textit{Planck}}
\newcommand{\wmap}{\textit{WMAP}}
\newcommand{\xmm}{\textit{XMM-Newton}}

\defcitealias{LeBrun2014}{L14}
\defcitealias{Arnaud2010}{A10}
\defcitealias{Planck2013}{PIntXI}

\title[Recovery of SZ flux of dark matter haloes]{Testing Sunyaev--Zel'dovich measurements of the hot gas content of dark matter haloes using synthetic skies}

\author[A.~M.~C.~Le Brun, I.~G.~McCarthy and J.-B.~Melin]{Amandine~M.~C.~Le Brun$^{1,2}$\thanks{E-mail: amandine.le-brun@cea.fr}, Ian~G.~McCarthy$^1$\thanks{E-mail: i.g.mccarthy@ljmu.ac.uk}, Jean-Baptiste~Melin$^3$\\
$^{1}$Astrophysics Research Institute, Liverpool John Moores University, 146 Brownlow Hill, Liverpool L3 5RF, United Kingdom\\
$^{2}$Laboratoire AIM, IRFU/Service d'Astrophysique -- CEA/DSM -- CNRS -- Universit\'e Paris Diderot, B\^at. 709, CEA-Saclay,\\ 91191 Gif-sur-Yvette Cedex, France\\
$^{3}$DSM/Irfu/SPP, CEA Saclay, 91191 Gif-sur-Yvette, France}

\begin{document}

\date{Accepted ... Received ...}

\pagerange{\pageref{firstpage}--\pageref{lastpage}} \pubyear{2015}

\maketitle

\label{firstpage}

\begin{abstract}
The thermal Sunyaev--Zel'dovich (tSZ) effect offers a means of probing the hot gas in and around massive galaxies and galaxy groups and clusters, which is thought to constitute a large fraction of the baryon content of the Universe. The \planck~collaboration recently performed a stacking analysis of a large sample of `locally brightest galaxies' (LBGs) and, surprisingly, inferred an approximately self-similar relation between the tSZ flux and halo mass. At face value, this implies that the hot gas mass fraction is {\it independent} of halo mass, a result which is in apparent conflict with resolved X-ray observations. We test the robustness of the inferred trend using synthetic tSZ maps generated from cosmological hydrodynamical simulations and using the same tools and assumptions applied in the \planck~study. We show that, while the detection and the estimate of the `total' flux (within $5 r_{500}$) is reasonably robust, the inferred flux originating from within $r_{500}$ (i.e.\ the limiting radius to which X-ray observations typically probe) is highly sensitive to the assumed pressure distribution of the gas. Using our most realistic simulations with AGN feedback, that reproduce a wide variety of X-ray and optical properties of groups and clusters, we estimate that the derived tSZ flux within $r_{500}$ is biased high by up to to an order of magnitude for haloes with masses $M_{500} \sim 10^{13}$ M$_{\odot}$. Moreover, we show that the AGN simulations are consistent with the total tSZ flux--mass relation observed with \planck, whereas a self-similar model is ruled out.
\end{abstract}

\section{Introduction}

Developing a detailed understanding of the formation and evolution of galaxies is one of the `holy grails' of modern astrophysics. It has become clear in recent years that achieving this goal requires developing physical models that faithfully capture the thermodynamic history of the gas in dark matter haloes, of which only a very small fraction is able to cool and collapse enough to form stars (e.g.\ \citealt{Balogh2001,Budzynski2014}). The problem is a challenging one, as the gas can be affected by a myriad of physical processes, such as feedback from supernovae and Active Galactic Nuclei (hereafter AGN), radiative cooling and associated thermal instabilities, as well as the action of magnetic fields, thermal conduction, viscosity, plasma instabilities, etc. The situation is made even more difficult by the fact that a large fraction of the baryons are `missing' observationally (e.g.\ \citealt{Fukugita1998}). That is, only a relatively small fraction of the total baryonic content of the Universe (as inferred by measurements of the cosmic microwave background, hereafter CMB, or implied by big bang nucleosynthesis theory combined with measurements of light elements at high redshift) has been mapped in any detail. Simple theoretical calculations, as well as the predictions of detailed hydrodynamical simulations, suggest that a large fraction of the missing baryons is in the form of warm-hot ($10^{5-7}$ K) gas within and surrounding dark matter haloes (e.g.\ \citealt{White1978,White1991,Cen1999,Dave2001}), but this gas has so far been very difficult to detect observationally (e.g.\ \citealt{Bregman2007}).

\pagebreak
X-ray observations offer one way to probe this component, but they are generally limited to relatively high-mass systems (clusters and X-ray-bright groups) and to only the central regions of nearby galaxies and low-mass groups, whereas the bulk of the gas/baryons are expected to be at much larger radius. The thermal Sunyaev--Zel'dovich (tSZ) effect \citep{Sunyaev1970,Sunyaev1972}, which is a spectral distortion of the CMB background due to inverse Compton scattering of CMB photons off hot gas, has recently emerged as a new tool for probing this hot gas component. Since the tSZ effect depends only linearly on the gas density (as opposed to X-ray emission, which depends on the square of the density) it is more sensitive to gas at large radii. Binning (or `stacking') large numbers of systems offers a means of probing the average hot gas content down to low mass. This technique has recently been exploited in a number of recent studies, such as the binning of optically-selected and X-ray-selected clusters using tSZ data from {\it WMAP} and \planck~(e.g.\ \citealt{Afshordi2007,Bielby2007,Melin2011,Planck2011a,Planck2011c}), as well as the binning of Sloan Digital Sky Survey (SDSS)-selected Luminous Red Galaxies using observations made with the Atacama Cosmology Telescope \citep{Hand2011}.

Recently, the \planck~collaboration reported the detection of significant amounts of hot gas via the tSZ effect down as far as $M_{500} \approx 4\times10^{12}$ M$_\odot$ (i.e.\ in systems only a factor a few more massive than the Milky Way and M31) by performing a binning analysis on a large sample of Locally Brightest Galaxies (LBGs) from the SDSS (\citeauthor{Planck2013}, hereafter \citetalias{Planck2013}; see also \citealt{Greco2014}). Intriguingly, the inferred mean relation between the integrated tSZ flux and halo mass is close to the self-similar expectation. That is, taken at face value, the results suggest that galaxies, groups, and clusters have nearly the same baryon fraction measured within $r_{500}$\footnote{$r_{500}$ is defined as the radius which encloses a mean density that is 500 times the critical density of the Universe at the object's redshift.}. This is a surprisingly result to say the least, as numerous X-ray studies of X-ray-bright galaxy groups and clusters have consistently found that groups show a marked deficit of baryons with respect to their more massive cousins \citep[e.g.][]{David2006,Gastaldello2007,Pratt2009,Sun2009}. Furthermore, cosmological hydrodynamical simulations that invoke AGN feedback and that can reproduce the X-ray gas mass fractions of groups and clusters without overcooling (e.g.\ \citealt{Puchwein2008,Fabjan2010,McCarthy2010,Battaglia2013,LeBrun2014}) predict that the gas mass fractions will continue to steadily decline well down into the individual galaxy regime. Therefore, if it holds, the \citetalias{Planck2013} result has fundamental implications for theories of galaxy formation and cosmic feedback.

However, the measurement of the tSZ signal down to such low masses is not a simple one by any means. Not only is stacking required, but strong assumptions about the distribution of the hot gas are required to estimate the flux within radii that are comparable to (or smaller than) the instrumental beam size. Furthermore, the contribution of both correlated and uncorrelated line-of-sight hot gas has not yet been well quantified. Fortunately, the impact of all these effects on the recovered signal can be tested using synthetic observations from a suite of reasonably realistic hydrodynamical simulations.

In the present study, we undertake a critical assessment of the recovery of the integrated tSZ flux ($Y_{500}$) as a function of halo mass in current observations, by using synthetic observations generated from a suite of state-of-the-art, large-volume cosmological hydrodynamical simulations (cosmo-OWLS). Although our methodology and tests have been geared towards a comparison with the trends reported recently in \citetalias{Planck2013}, the conclusions should be broadly applicable to stacked tSZ flux measurements based on matched filter techniques.

The paper is organized as follows. In Section 2, we briefly describe the hydrodynamical simulations (Section 2.1), our tSZ mapmaking techniques (Section 2.2), and our halo selection and flux recovery approach (Section 2.3). In Section 3, we compare the true and recovered tSZ flux--mass relations and quantify the difference (bias) between the two. In Section 4, we re-examine the pressure profiles of the hot gas in the simulations and design a new mass-dependent spatial template for the tSZ matched filter. Finally, in Section 5, we summarize and discuss our findings. 

\section{Simulations}
\label{sec:sims}

\subsection{cosmo-OWLS}
\label{sec:owls}

\begin{table*}
\centering
\begin{tabular}{|l|l|l|l|l|l|l|}
   \hline
	Simulation & UV/X-ray background & Cooling & Star formation & SN feedback & AGN feedback & $\Delta T_{heat}$ \\
	\hline
  \nocool & Yes & No & No & No & No & ...\\
  \refsim & Yes & Yes & Yes & Yes & No & ...\\
  \agn~8.0 & Yes & Yes & Yes & Yes & Yes & $10^{8.0}$ K\\
  \agn~8.5 & Yes & Yes & Yes & Yes & Yes & $10^{8.5}$ K\\
  \hline
\end{tabular}
\caption{cosmo-OWLS runs used here and their included sub-grid physics.}
\label{table:owls}
\end{table*}

We use the cosmo-OWLS suite of cosmological hydrodynamical simulations described in detail in \citet{LeBrun2014} \citep[hereafter L14; see also][]{McCarthy2014,vanDaalen2014,Velliscig2014}. They form an extension to the OverWhelmingly Large Simulations project \citep[OWLS;][]{Schaye2010}. The cosmo-OWLS suite consists of large volume ($400~h^{-1}$ comoving Mpc on a side) periodic box simulations with $1024^3$ dark matter and $1024^3$ baryonic particles with updated initial conditions derived either from the 7-year {\it Wilkinson Microwave Anisotropy Probe} (\wmap) data \citep{Komatsu2011} \{$\Omega_{m}$, $\Omega_{b}$, $\Omega_{\Lambda}$, $\sigma_{8}$, $n_{s}$, $h$\} = \{0.272, 0.0455, 0.728, 0.81, 0.967, 0.704\} or the \planck~data (\citeauthor{Planck_cosmology}) = \{0.3175, 0.0490, 0.6825, 0.834, 0.9624, 0.6711\}. This yields dark matter and (initial) baryon particle masses of $\approx4.44\times10^{9}~h^{-1}~\textrm{M}_{\odot}$ ($\approx3.75\times10^{9}~h^{-1}~\textrm{M}_{\odot}$) and $\approx8.12\times10^{8}~h^{-1}~\textrm{M}_{\odot}$ ($\approx7.54\times10^{8}~h^{-1}~\textrm{M}_{\odot}$), respectively for the \planck~(\wmap7) cosmology. The gravitational softening is set to $4~h^{-1}$ kpc (in physical coordinates below $z=3$ and in comoving coordinates at higher redshifts). Note that we use $N_{ngb}=48$ neighbours for the smoothed particle hydrodynamics (SPH) interpolation and the minimum SPH smoothing length is set to one tenth of the gravitational softening. Below we present results based on the \wmap7 runs only, but the results and conclusions are insensitive to the choice of cosmology.

In addition to these large-volume runs, we have carried out additional runs with eight (two) times higher mass (spatial) resolution but in smaller volumes (the boxes are $100~h^{-1}$ comoving Mpc on a side). The motivation for this is two-fold: (i) to allow an exploration of the role of numerical convergence on the results; and (ii) to test the recovery of the tSZ flux down to lower halo masses than can be reliably probed with our larger, lower resolution boxes. Note that the Locally Brightest Galaxy (LBG) sample of \citetalias{Planck2013} spans approximately three orders of magnitude in halo mass, from massive clusters with $M_{500}\sim2-3\times10^{15}~\textrm{M}_{\odot}$ down to individual galaxies with $M_{500}\sim2-3\times10^{12}~\textrm{M}_{\odot}$.

The simulations were carried out with a version of the Lagrangian TreePM-SPH code \textsc{gadget3} \citep[last described in][]{Springel2005a}, which has been significantly modified to include new `sub-grid' physics. Starting from identical initial conditions (in each cosmology), the key parameters that govern the most important sub-grid physics, including feedback from supernovae (SNe) and active galactic nuclei (AGN), are systematically varied. We use four of the five physical models described in \citetalias{LeBrun2014}: a non-radiative model (\nocool); a model (\refsim~which corresponds to the OWLS reference model) which includes prescriptions for metal-dependent radiative cooling \citep*{Wiersma2009a}, stellar evolution, mass loss and chemical enrichment \citep{Wiersma2009b}, star formation \citep{Schaye2008} and kinetic stellar feedback \citep{DallaVecchia2008} and two models (\agn~8.0, which was simply called \agn~in the original OWLS papers, and \agn~8.5) which further include a prescription for supermassive black hole growth and AGN feedback \citep{Booth2009}, which is a modified version of the model developed by \citet*{Springel2005b}. The black holes store the feedback energy until they can heat neighbouring gas particles by a pre-determined amount $\Delta T_{heat}$. As in \citet{Booth2009}, 1.5 per cent of the rest-mass energy of the gas which is accreted on to the supermassive black holes is used for the feedback. This results in a satisfactory match to the normalization of the black hole scaling relations (\citealt{Booth2009}; see also \citetalias{LeBrun2014}) which is independent of the exact value of $\Delta T_{heat}$. The two \agn~models used here only differ by their value of $\Delta T_{heat}$, which is the most crucial parameter of the AGN feedback model in terms of the hot gas properties of the resulting simulated population of groups and clusters (\citealt{McCarthy2011}; \citetalias{LeBrun2014}). It is set to $\Delta T_{heat}=10^{8}$ K for \agn~8.0 and  $\Delta T_{heat}=3\times10^{8}$ K for \agn~8.5. Note that since the same quantity of gas is being heated in the \agn~8.5 model as in the \agn~8.0 model, more time is needed for the black holes to accrete enough gas to heat the surrounding gas to a higher temperature. Increasing the heating temperature hence results into more bursty and more energetic feedback episodes.

Table~\ref{table:owls} provides a list of the runs used here and the sub-grid physics that they include.

These models have been compared to a wide range of observational data by both \citetalias{LeBrun2014} and \citet{McCarthy2014}. In \citetalias{LeBrun2014}, we focused on the comparison to the resolved hot gas (e.g.\ X-ray luminosities and temperatures, gas fraction, entropy and density profiles, integrated tSZ flux) and stellar properties (e.g.\ $I$-band total-mass-to-light ratio, dominance of the brightest cluster galaxies) of local galaxy groups and clusters, as well as the properties of the central black hole and concluded that the fiducial AGN model (\agn~8.0) produces a realistic population of galaxy groups and clusters, broadly reproducing both the median trend and, for the first time, the scatter in physical properties over approximately two decades in mass ($10^{13}~\textrm{M}_{\odot} \la M_{500} \la 10^{15}~\textrm{M}_{\odot}$) and 1.5 decades in radius ($0.05 \la r/r_{500} \la 1.5$). In \citet{McCarthy2014}, we explored the sensitivity of the thermal Sunyaev--Zel'dovich power spectrum to important non-gravitational physics and also showed that the fiducial AGN model adequately matches the observed pressure profiles of local groups and clusters (see their fig. 2).

\subsection{Theoretical thermal Sunyaev--Zel'dovich (tSZ) maps}
\label{sec:ttSZmaps}

The thermal tSZ signal is characterized by the dimensionless Compton $y$ parameter, defined as:
\begin{equation}
y\equiv\frac{\sigma_T}{m_e c^2}\int P_edl
\end{equation}
where $\sigma_T$ is the Thomson cross-section, $c$ the speed of light, $m_e$ the electron rest-mass and $P_e=n_ek_BT_e$ is the electron pressure with $k_B$ being the Boltzmann constant. The integration is done along the observer's line of sight.

Compton $y$ maps are generated by stacking randomly transformed (by a combination of translations, rotations and axis inversions) snapshots along the observer's line of sight (e.g.\ \citealt{daSilva2000}). The light cones extend back to $z=3$. Ten (sixteen) quasi-independent realisations are generated for the large lower resolution (smaller higher resolution) simulations by randomly varying the light cone transformations. As the methods used for the production of the Compton $y$ maps are described in some detail in \citet{McCarthy2014}, we will only present a brief summary below.

We follow the method of \citet{Roncarelli2006,Roncarelli2007} and compute 
\begin{equation}
\Upsilon_i\equiv\frac{\sigma_T}{m_e c^2}\frac{k_BT_im_i}{\mu_{e,i}m_H}
\end{equation}
for the $i^{th}$ gas particle, where $T_i$, $m_i$ and $\mu_{e,i}$ are respectively the temperature, mass and mean molecular weight per free electron of the gas particle and $m_H$ is the atomic mass of hydrogen. The contribution to the Compton $y$ parameter by the $i^{th}$ particle is given by 
\begin{equation}
y_{i}\equiv\Upsilon_i/L^2_{pix,i},
\end{equation}
where $L^2_{pix,i}$ is the physical area of the pixel in which the $i^{th}$ particle falls at the angular diameter distance from the observer to the particle. Finally, $y_i$ is SPH-smoothed onto the map using the SPH smoothing kernel which was used by \textsc{gadget}3 for the computation of the hydrodynamical forces and the three-dimensional smoothing length of the particle expressed in angular units (i.e.\ the smoothing length divided by the distance to the particle from the observer). 

We produce maps that are five degrees (1.25 degrees) on a side for the large lower resolution (smaller higher resolution) simulations. This roughly corresponds to the angular size of the $400~h^{-1}$ comoving Mpc ($100~h^{-1}$ comoving Mpc) box at $z=3.0$. The maps have an angular pixel size of 2.5 arcseconds.

As discussed in \citet{McCarthy2014}, because the realisations are produced using the same simulations, they are only quasi-independent. At high redshift, the light cone samples most of the volume of simulation and thus the different realisations contain many of the same structures (though at randomised locations). At lower redshift, however, the cones sample only a small fraction of the simulated volume and the different realisations are therefore effectively independent. This is relevant for the present study, which is focused on the recovery of the tSZ flux of relatively local ($z \le 0.4$) dark matter haloes. The rationale behind using several quasi-independent realisations is to mitigate the impact of cosmic variance by significantly increasing the sample size.

\subsection{Halo selection and tSZ flux recovery}
\label{sec:mockLBG}

The tSZ signal of {\it individual} galaxies, groups, and low-mass clusters cannot normally be detected with \planck. Large numbers of systems must therefore be binned (`stacked') to measure the mean relation between haloes and tSZ flux down to low masses. \citetalias{Planck2013} undertook such a binning analysis of some $\sim260~000$ locally brightest galaxies selected from the SDSS. They defined their locally brightest galaxy (hereafter LBG) sample as all the galaxies with $z>0.03$ and $r < 17.7$ which are brighter than any other galaxy within a projected distance of 1.0 Mpc and with a redshift difference smaller than $1,000~\textrm{km s}^{-1}$. The sample was derived from the spectroscopic New York University Value Added Galaxy Catalogue (NYU-VAGC)\footnote{http://sdss.physics.nyu.edu/vagc/}, which is based on the seventh data release of the Sloan Digital Sky Survey \citep[SDSS/DR7;][]{Abazajian2009} and covers 7966 square degrees. The NYU-VAGC also provides, among other properties, stellar masses which were computed by \citet{Blanton2007} by fitting stellar populations to the five-band SDSS photometry assuming a \citet{Chabrier2003} initial mass function. 

\citetalias{Planck2013} derived the mean tSZ flux in bins of stellar mass. To deduce the more fundamental relation between tSZ flux and halo mass, they used the semi-analytic galaxy formation model of \citet{Guo2011}, which was tuned to closely match the observed luminosity and stellar mass functions of SDSS galaxies in a \wmap7 cosmology \citep[][]{Guo2013}. When deriving the relation between tSZ flux and halo mass, the effects of scatter in the stellar mass--halo mass relation, as well as (minor) contamination of the LBG sample by satellite systems, are accounted for using the semi-analytic model.

Our aim is to test the accuracy of the recovered tSZ flux--halo mass relation by analysing synthetic maps in a way that is faithful to that done for the real data. We place our emphasis on the recovery of the tSZ flux, rather than testing the conversion of the tSZ flux--stellar mass relation into an tSZ flux--halo mass relation.  The latter will depend on the models correctly capturing the intrinsic scatter in the stellar mass--halo mass relation (i.e.\ in any given stellar mass bin, there will be a range of halo masses and since the SZ flux has a steeper than linear scaling with halo mass, the mean signal will be driven by the higher halo mass systems in the bin).  The scatter can be constrained to some degree by requiring the models to match, for example, the galaxy stellar mass function and/or the clustering of galaxies in bins of stellar mass (as is the case for the model of \citealt{Guo2011}, used in the {\it Planck} LBG study), but it is probably fair to say that the intrinsic scatter issue is still under consideration.  This uncertainty in the stellar mass--halo mass relation could therefore potentially further complicate the physical interpretation of the results presented in \citetalias{Planck2013}.

We first construct halo catalogues corresponding to the maps described in Section~\ref{sec:ttSZmaps} using a standard friends-of-friends (FoF) algorithm run on the snapshot data. The catalogues contain the positions on the map, the angular size $\theta_{500}$ ($\equiv r_{500}/d_A$, where $d_A$ is the angular diameter distance to the halo) and the halo mass $M_{500}$ of all the haloes with $z<0.4$ and $M_{500}>2\times10^{13}~\textrm{M}_\odot$ ($M_{500}>10^{12}~\textrm{M}_\odot$) for the large lower resolution (smaller higher resolution) simulations. The redshift and mass thresholds have been chosen to roughly reproduce the bounds of the \citetalias{Planck2013} LBG sample. To produce `dirty' maps with similar characteristics to the real data, the raw maps are downgraded from their original 2.5 arcseconds resolution to a 0.83 arcminute resolution and the effects of primary CMB, of the \planck~beams and their associated noise are added to obtain six synthetic observations at the frequencies of the \planck~HFI instrument (100, 143, 217, 353, 545 and 857 GHz). Note that in producing the dirty maps, we have neglected dust and radio point source contamination, which can be significant for real observations. However, this omission is at least partially compensated for by the fact that the overlapping SDSS-\planck~survey area (7966 square degrees) greatly exceeds the survey area that can be simulated with current self-consistent cosmological hydro simulations --- the cosmo-OWLS maps, which are some of the largest ever produced, are still only 25 square degrees each. In practice, we produce as many (primary CMB+instrumental) noise realisations of the simulation maps as are needed to obtain statistical error bars on the derived tSZ flux--halo mass relation that are comparable to the ones reported in \citetalias{Planck2013}, namely two (twenty-five) for the large lower resolution (smaller higher resolution) simulations.

Following the production of the dirty maps, we then apply the same multi-frequency matched filter \citep[][MMF]{Herranz2002,Melin2006} that was used by the \planck~collaboration. The MMF yields an estimate of the tSZ flux within $5r_{500}$, which is then converted using a constant conversion factor into the tSZ flux within $r_{500}$ (see discussion below), as characterized by the value of its spherically integrated Compton parameter within $r_{500}$:
\begin{equation}
d_{A}(z)^{2}Y_{500}=\frac{\sigma_T}{m_ec^2}\int P_edV
\end{equation}
where $d_A(z)$ is the angular diameter distance and the integration is done over a sphere of radius $r_{500}$. Throughout the paper, unless otherwise stated, we use the quantity 
\begin{equation}
\tilde{Y}_{500}\equiv Y_{500}E^{-2/3}(z)\left(\frac{d_A(z)}{500~\textrm{Mpc}}\right)^{2}
\end{equation}
where $E(z)\equiv H(z)/H_0=\sqrt{\Omega_m(1+z)^3+\Omega_\Lambda}$ gives the redshift evolution of the Hubble parameter $H(z)$ in a flat $\Lambda$CDM Universe. $\tilde{Y}_{500}$ corresponds to the intrinsic tSZ flux self-similarly scaled to $z=0$ and scaled to a fixed angular diameter distance. 

The MMF is optimised in both frequency and angular space\footnote{It maximises the signal-to-noise ratio of objects which follow the assumed spectral and spatial templates.} by assuming the known frequency dependence of the thermal tSZ effect and the `universal pressure profile' \citep[][hereafter \citetalias{Arnaud2010}]{Arnaud2010}, derived from a combination of X-ray observations of the \xmm~REXCESS cluster sample \citep{Bohringer2007} at radii of $r \la r_{500}$ and hydrodynamical simulations at radii of $r_{500} \la r \la 5 r_{500}$. Consistent with \citetalias{Planck2013}, the MMF is run in a non-blind mode using the positions and sizes, $\theta_{500}$, from the halo catalogue. The MMF then gives a measure of the strength of the tSZ signal $\tilde Y_{500}(i)$ and its associated measurement uncertainty $\tilde\sigma_{\theta_{500}}(i)$ for the halo surrounding the $i^{th}$ galaxy. The measurement uncertainty takes into account the statistical uncertainties due to astrophysical (e.g.\ primary CMB) and instrumental noise, but not the uncertainties due to halo modelling (e.g.\ shape of the pressure profile, size). Note that the tSZ spectral function was not integrated over the \planck~bandpasses for each frequency for both constructing the `dirty' maps and creating the matched filter for each of the \planck~frequencies. This should have no impact upon the results as it was done at both stages.

It is important to note here that the LBGs are at best only marginally resolved by \planck. Therefore, in practice, the tSZ flux is actually measured within the larger aperture of $5\theta_{500}$;
\begin{equation}
Y_{5 r_{500}} = \int_0^{5\theta_{500}} 2\pi \theta y(\theta) d\theta 
\end{equation}
The flux within the spherical aperture $r_{500}$ is then computed assuming the spatial template used in the MMF (the universal pressure profile), leading to a conversion factor $Y_{500}=Y_{5r_{500}}/1.796$. (It is assumed that there is no flux originating from beyond $5r_{500}$ and therefore $Y_{5\theta_{500}} = Y_{5 r_{500}}$.)  It is perhaps an obvious statement that if in reality the adopted spatial template poorly describes the pressure distribution of the hot gas, this will lead to a systematic error in the recovered flux within $r_{500}$. We will return to this point below.

Finally, for both simulations and observations, the tSZ flux is binned by mass with the bin-average flux and the corresponding uncertainty given by (\citeauthor{Planck2011a,Planck2011c}; \citetalias{Planck2013})
\begin{equation}
\langle \tilde Y_{500}\rangle_b=\frac{\sum_{i=1}^{N_b} \tilde Y_{500}(i)/\tilde\sigma^2_{\theta_{500}}(i)}{\sum_{i=1}^{N_b} 1/\tilde\sigma^2_{\theta_{500}}(i)}
\label{eq:Yweighted}
\end{equation} 
and
\begin{equation}
\sigma_b^{-2}=\sum_{i=1}^{N_b} 1/\tilde\sigma^2_{\theta_{500}}(i),
\end{equation}
where $N_b$ is the number of galaxies in bin $b$. The integrated tSZ flux is often normalized by the self-similar integrated tSZ flux--mass relation, as given in appendix B of \citetalias{Arnaud2010}:
\begin{equation}
\tilde Y_{500,A10}=9.07\times10^{-4}\left[\frac{M_{500}}{3\times10^{14}h_{70}^{-1}\textrm{M}_\odot}\right]^{5/3}h_{70}^{-1}~\textrm{arcmin}^2
\label{eq:Y500A10}
\end{equation}
 
\section{The need for synthetic tSZ observations}
\label{sec:synthetic}

\begin{figure}
\begin{center}
\includegraphics[width=1.0\hsize]{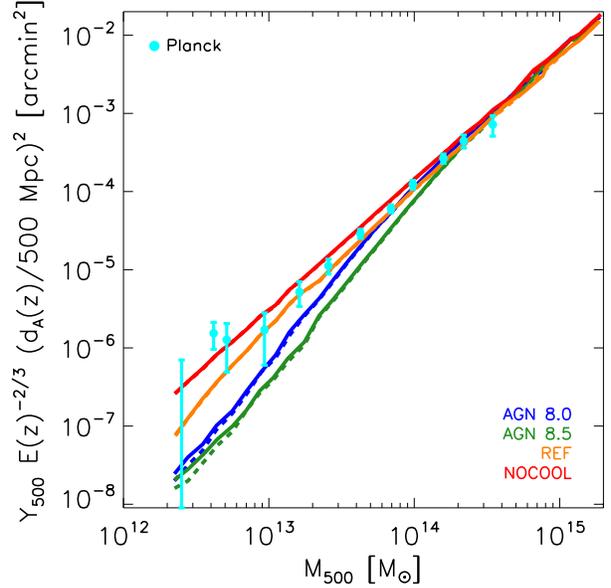}
\caption{Comparison of the $Y_{500}$--$M_{500}$ relation inferred by \citetalias{Planck2013} with the intrinsic (true) $z=0$ relation from the cosmo-OWLS runs. The filled cyan circles with error bars represent the observational data of \citetalias{Planck2013}. The solid and dashed curves (red, orange, blue and green) represent the mean and median integrated tSZ flux--$M_{500}$ relations in bins of $M_{500}$ for the different simulations, respectively. Taken at face value, the \citetalias{Planck2013} results favour a close to self-similar $Y_{500}$--$M_{500}$ relation. However, $Y_{500}$ is not directly measured and synthetic observations are therefore required for proper comparison to the inferred trend (see Fig.~\ref{fig:mmfcomparison}).}
\label{fig:Y500_M500}
\end{center}
\end{figure}

\begin{figure*}
\begin{center}
\includegraphics[width=0.4898\hsize]{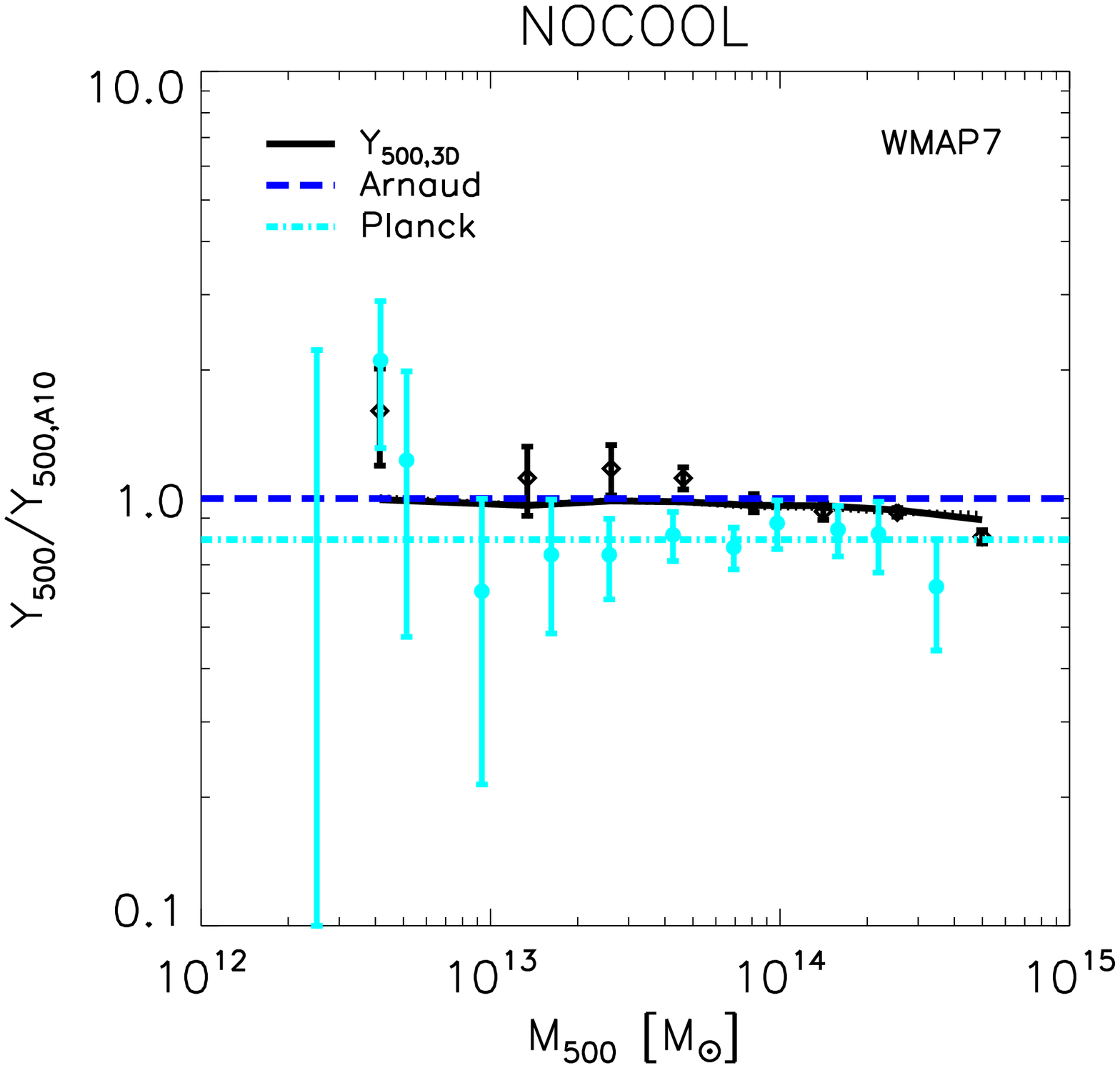}
\includegraphics[width=0.4898\hsize]{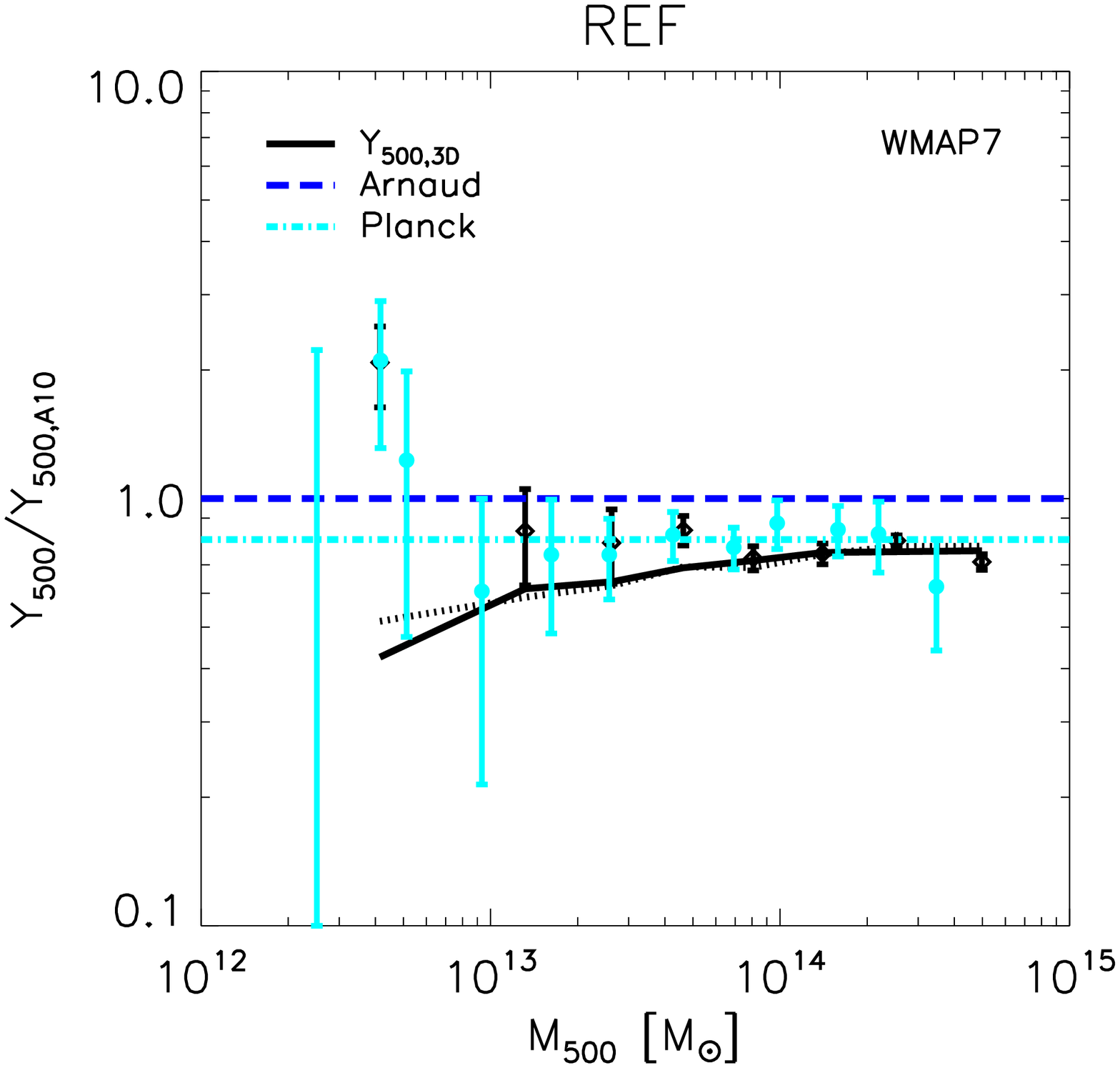}
\includegraphics[width=0.4898\hsize]{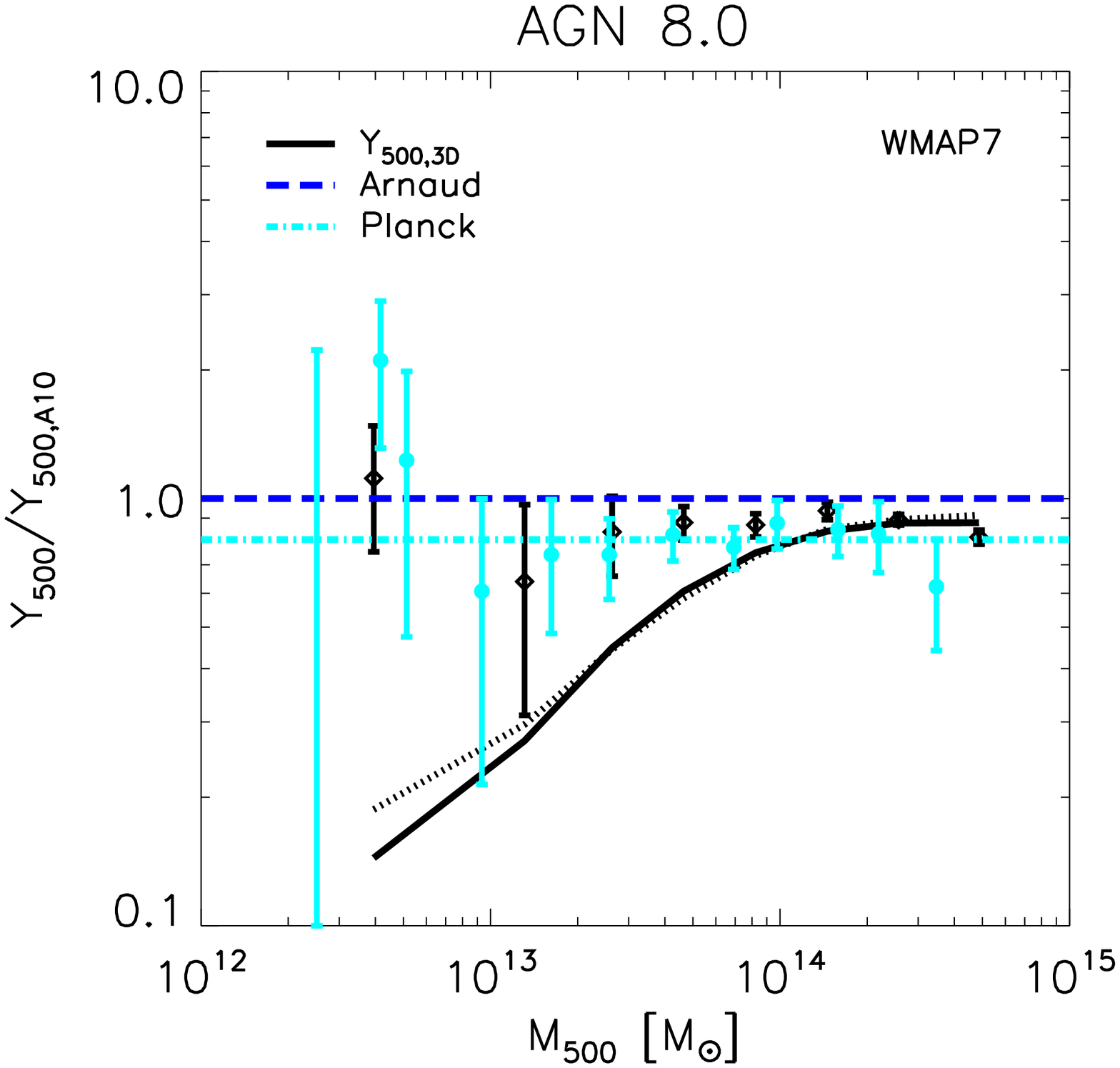}
\includegraphics[width=0.4898\hsize]{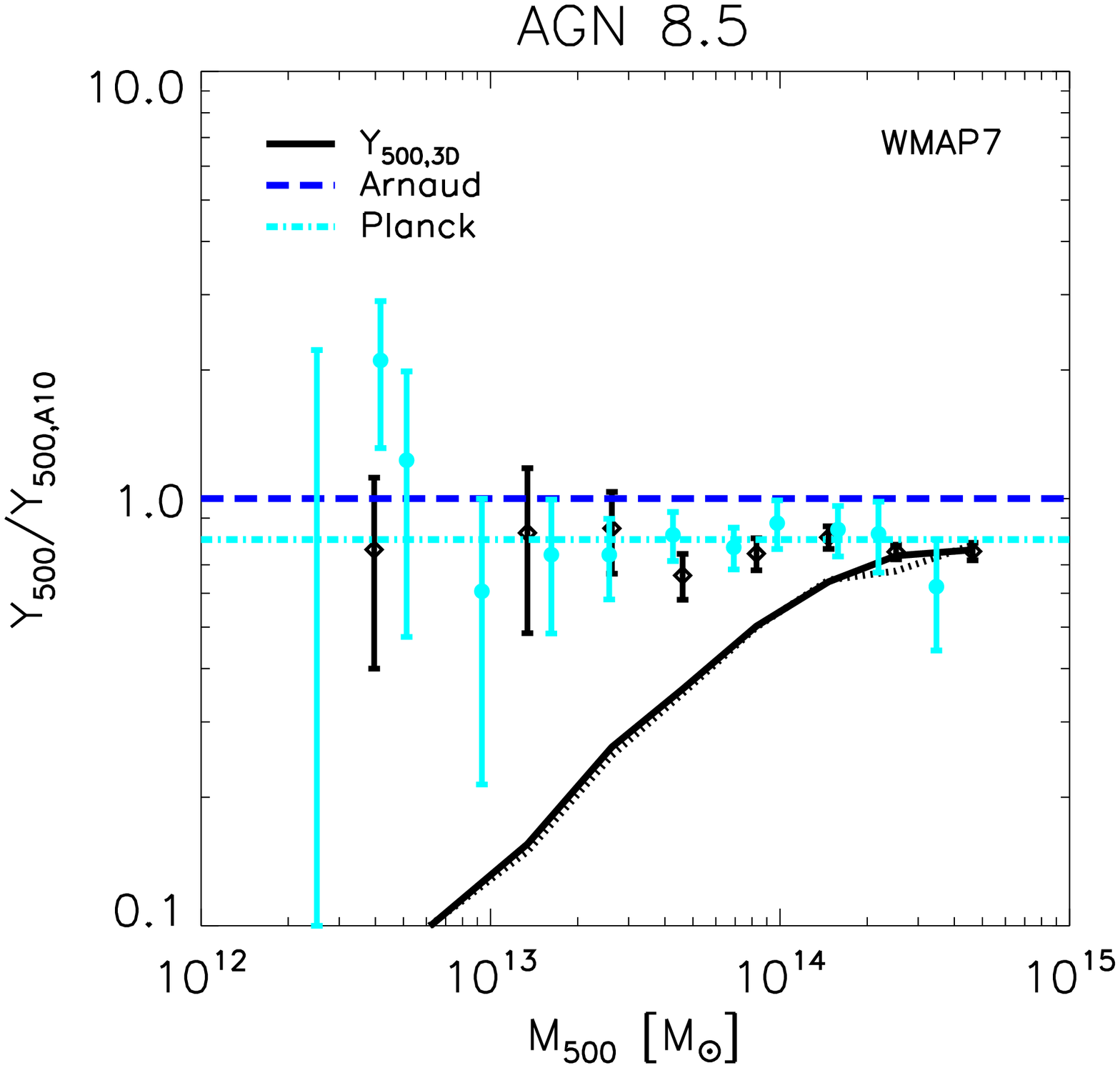}
\caption{Comparison of the $Y_{500}$--$M_{500}$ relation inferred by \citetalias{Planck2013} with the recovered relation from synthetic tSZ observations of the cosmo-OWLS runs. The blue dashed and cyan dot-dashed lines correspond to the best-fitting scaling relations of \citetalias{Arnaud2010} and \citetalias{Planck2013}, respectively. The solid and dotted black curves correspond to the unweighted and weighted mean {\it true} tSZ flux--$M_{500}$ relations in bins of $M_{500}$ for the different simulations, respectively. The filled cyan circles and empty black diamonds with error bars correspond to the observational data of \citetalias{Planck2013} and the results of the synthetic observations of the simulated maps, respectively. The integrated tSZ fluxes have been normalized using the best-fitting scaling relation of \citetalias{Arnaud2010}. The different panels correspond to the different physical models. The MMF recovers the intrinsic $Y_{500}$--$M_{500}$ well for models that neglect efficient feedback (\nocool~and \refsim). However, a significant bias is present in the recovered relation for models which include AGN feedback. The bias increases strongly with decreasing halo mass and with increasing AGN heating temperature (i.e.\ more violent feedback events).}
\label{fig:mmfcomparison}
\end{center}
\end{figure*}

We begin by comparing, in  Fig.~\ref{fig:Y500_M500}, the \citetalias{Planck2013} results to the {\it true} spherically integrated tSZ flux--mass ($Y_{500}-M_{500}$) relation of the simulated galaxy, group and cluster populations for the four physical models at $z=0$. The solid and dashed curves respectively represent the mean and median relations in bins of $M_{500}$ for the different simulations. The high-resolution simulations are used below $M_{500}=2\times10^{13}~\textrm{M}_{\odot}$ and the production runs above that threshold (see Appendix A). The filled cyan circles with error bars represent the observational data of \citetalias{Planck2013}.

Focusing first on the simulations, there is a visible steepening of the $Y_{500}$--$M_{500}$ relation in the galaxy and group regime ($M_{500} \la 10^{14}$ M$_\odot$) with the inclusion of AGN feedback. This is due to the fact that the AGN lower the gas fractions, and hence the integrated Compton y values, of low-mass haloes (e.g.\ \citealt{Puchwein2008,McCarthy2010}; L14) by ejecting gas from their high-redshift progenitors \citep{McCarthy2011}.

Interestingly, it is apparent that, taken at face value, the \citetalias{Planck2013} results appear to favour an approximately self-similar $Y_{500}$--$M_{500}$ relation, as obtained in non-radiative simulations (such as \nocool; red lines). This is consistent with the conclusions of \citetalias{Planck2013}. As already discussed in Section 1, this is a surprising result, given that numerous previous studies have demonstrated that the self-similar model is strongly ruled out by X-ray observations. In L14 (see also \citealt{McCarthy2010}), for example, we showed that the fiducial AGN model (\agn~8.0) reproduces a wide variety of X-ray and optical measurements of local groups and clusters (including their gas fractions) while the \nocool~model is strongly disfavoured.

As discussed in Section~\ref{sec:mockLBG}, however, the quantity $Y_{500}$ is not directly measured with \planck. Instead, what is measured is an estimate of the integrated flux within a 2D aperture $5 \theta_{500}$, and a scaling factor which depends on the assumed distribution of the hot gas must be applied to estimate the tSZ flux within the spherical radius $r_{500}$. Furthermore, hot gas along the line-of-sight can potentially bias the 2D flux. The magnitude of these effects can be quantified using synthetic observations of the type described in Section 2.3.

With this in mind, in Fig.~\ref{fig:mmfcomparison}, we compare the recovered tSZ flux--$M_{500}$ relation from our synthetic (`dirty') maps (empty black diamonds) with the mean true tSZ flux--$M_{500}$ relation, both unweighted (solid lines) and weighted (dotted lines; as given by equation~\ref{eq:Yweighted}), as well as with the observational data from \citetalias{Planck2013} (filled cyan circles with error bars). To more clearly emphasize the level of agreement (or lack thereof) between the recovered and true relations and the observational data, we normalize the tSZ flux using the best-fitting scaling relation of \citetalias{Arnaud2010} (see equation~\ref{eq:Y500A10}). The different panels correspond to the different physical models: from \nocool~(top left) to \agn~8.5 (bottom right) through \refsim~(top right) and \agn~8.0 (bottom left). 

Focusing first on the top two panels of Fig.~\ref{fig:mmfcomparison}, corresponding to the \nocool~and \refsim~simulations, we find that the MMF recovers the true relations relatively well, as deduced by comparing the empty black diamonds with the black curves. (The only exception to this is the lowest-mass bin, where the recovered flux is significantly higher than the intrinsic one, which we discuss below in more detail.)  It is interesting to note that the recovery is best for the \nocool~model, which has a nearly self-similar behaviour, in the sense that the shape of the pressure distribution of the hot gas is nearly independent of halo mass. It is worth re-iterating that the (empirical) universal pressure profile of A10 was derived from a sample of very massive clusters, where the high mass basically ensures that non-gravitational physics (e.g.\ AGN) has a relatively minor effect on the tSZ flux. Thus, the fact that the MMF recovers the true relation well for a simulation where the gas approximately follows the universal pressure profile on all mass scales may not be that surprising. In fact, it is reassuring.

Moving on to the bottom two panels of Fig.~\ref{fig:mmfcomparison}, corresponding to the two AGN models, we find a significant offset between the recovered and true $Y_{500}$--$M_{500}$ relations. The amplitude of the bias increases strongly with decreasing halo mass and with increasing AGN heating temperature. Notably, {\it the bias reaches nearly an order of magnitude at the lowest halo masses}. The fact that the models with AGN much more closely reproduce the X-ray and optical properties of real groups and clusters than the \nocool~and \refsim~simulations provides a strong impetus to understand the origin of the bias and to offer a means for correcting for it.

\section{Role of Deviations from the universal pressure profile}
\label{sec:nonUPP}

\begin{table*}
\centering
\begin{tabular}{llcccccccc}
\hline
Simulation & Median or mass-weighted & $P_{0,0}$ & $\alpha$ & $\beta$ &
$\gamma$ & $c_{500,0}$ & $\delta$ & $\epsilon$  \\
\hline
\refsim     & Mass-weighted & \phantom{5000}0.528 & 2.208 & 3.632 &
\phantom{$-$}1.486 & 1.192 & \phantom{$-$}0.051 & \phantom{$-$}0.210 \\
\refsim     & Median               & \phantom{5000}0.694 & 1.489 & 4.512
& \phantom{$-$}1.174 & 0.986 & \phantom{$-$}0.072 & \phantom{$-$}0.245 \\
\agn~8.0 & Mass-weighted  & \phantom{5000}0.581 & 2.017 & 3.835 &
\phantom{$-$}1.076 & 1.035 & \phantom{$-$}0.273 & \phantom{$-$}0.819 \\
\agn~8.0 & Median           & \phantom{5000}0.791 & 1.517 & 4.625 &
\phantom{$-$}0.814 & 0.892 & \phantom{$-$}0.263 & \phantom{$-$}0.805 \\
\agn~8.5 & Mass-weighted  & \phantom{5000}0.214 & 1.868 & 4.117 &
\phantom{$-$}1.063 & 0.682 & \phantom{$-$}0.245 & \phantom{$-$}0.839 \\
\agn~8.5 & Median           & \phantom{5000}0.235 & 1.572 & 4.850 &
\phantom{$-$}0.920 & 0.597 & \phantom{$-$}0.246 & \phantom{$-$}0.864 \\
\hline
\end{tabular}
\caption{Results of the pressure profile fitting.}
\label{table:Pfit}
\end{table*}

\begin{figure*}
\begin{center}
\includegraphics[scale=0.645]{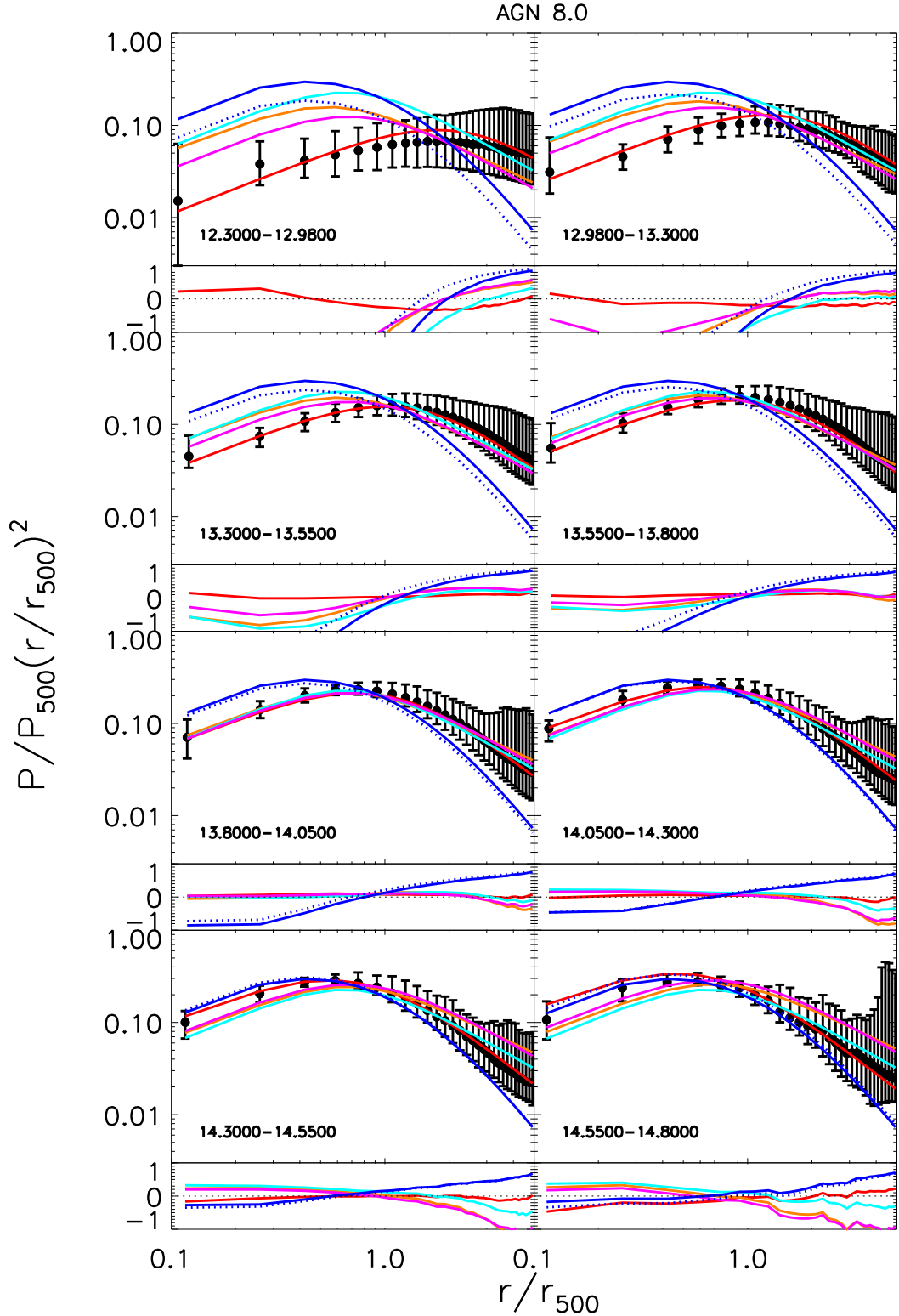}
\caption{Mass-weighted pressure profiles at $z=0$ for the \agn~8.0 model. The eight panels correspond to eight different mass bins (increasing from \emph{top left} to \emph{bottom right}). The bottom part of each panel shows the residuals of the best-fitting profiles. The black filled circles with error bars correspond to the median profile in the corresponding mass bin and the error bars encompass 68 per cent of the systems. The solid blue, dashed blue, solid cyan, orange, magenta and red lines correspond to the universal pressure profile of the appendix B of \citetalias{Arnaud2010}, the empirical universal pressure profile of section 5 of \citetalias{Arnaud2010}, the best-fitting GNFW functional form with all the parameters left free to vary, the best-fitting GNFW form but with the concentration now a power-law of mass, the best-fitting GNFW form but with the normalization now a power-law of mass, and the best-fitting GNFW form but with the concentration  and normalization now power-laws of mass, respectively. The shape of the pressure profiles is quite strongly mass-dependent and two of the GNFW coefficients must be made mass-dependent in order to obtain a reasonable fit (solid red curve) over the whole radial and mass ranges when AGN feedback is included.}
\label{fig:Pmwfit}
\end{center}
\end{figure*}

As the magnitude of the bias between the recovered and true $Y_{500}$--$M_{500}$ relations in Fig.~\ref{fig:mmfcomparison} evidently depends strongly the nature of non-gravitational (sub-grid) physics included in the simulations, this suggests its origin is tied to deviations of the hot gas distribution in the simulations from that adopted in the matched filter. We therefore explore this possibility in detail below. An alternative source of bias at the low-mass end is from source confusion (i.e.\ hot gas along the line of sight). We explore the role of source confusion in Appendix B, demonstrating that, while it is expected to significantly increase the scatter in the recovered fluxes of individual haloes, source confusion does not significantly bias the recovered {\it mean} tSZ flux--halo mass relation.

The `universal pressure profile' of A10 is described well by a so-called generalized NFW (GNFW) model and was first applied by \citet{Nagai2007} to describe the pressure distribution of the hot gas in clusters. It was subsequently shown to provide a relatively good description to large samples of (high-mass) X-ray groups and clusters (e.g.\ \citealt{Mroczkowski2009,Arnaud2010,Plagge2010,Sun2011}; \citeauthor{Planck_prof}; \citealt{Sayers2013,McDonald2014}). The model has five free parameters:
\begin{equation}
\frac{P(r)}{P_{500}}=\frac{P_0}{(c_{500}r/r_{500})^\gamma[1+(c_{500}r/r_{500})^\alpha]^{(\beta-\gamma)/\alpha}},
\label{eq:GNFW}
\end{equation}
where 
\begin{eqnarray}
P_{500} & \equiv & n_{e,500}k_BT_{500} \\
& = & \frac{500 f_b \rho_{crit}(z) \mu m_p GM_{500}}{\mu_e m_H 2r_{500}}, 
\end{eqnarray}
\noindent $f_b\equiv\Omega_b/\Omega_m$ is the universal baryon fraction, $\mu$ and $\mu_e$ are the mean molecular weight and the mean molecular weight per free electron, respectively. 

A10 fitted the generalised NFW profile to the REXCESS sample of X-ray clusters to constrain four of the five parameters ($P_0$, $c_{500}$, $\gamma$, and $\alpha$). Following \citet{Nagai2007}, A10 fixed the value of $\beta$ (the external slope) to $5.49$, the value which best describes the large-radii behaviour of the simulations of \citet{Nagai2007}. We note here that the pressure distribution of the gas as inferred by X-ray observations is only constrained out to $r \sim r_{500}$. Hydrodynamical simulations have therefore been relied on by previous studies to constrain the outer slope. However, recently resolved tSZ observations of local clusters with \planck~have now allowed for direct measurements of the pressure profile out to several $r_{500}$ and indicate that the external slope may be somewhat flatter ($\beta \approx 4$) than predicted by some previous simulations (\citeauthor{Planck_prof}). Interestingly, our cosmo-OWLS models predict outer pressure distributions that are in very good agreement with \citeauthor{Planck_prof} (see Fig.~\ref{fig:Pmwfit} below, comparing the simulated profiles with the universal pressure profile assuming $\beta = 5.49$).  One possible reason for the difference in pressure profiles between cosmo-OWLS and the simulations used by A10 is that the previous simulations lacked feedback from AGN, which is crucial for preventing overcooling, which should lead to reduced pressures.  \citet{Battaglia2012}, who use the AGN simulations of \citet{Battaglia2010}, have come to similar conclusions.

We note that \citet{Nagai2007} justified the choice of this particular functional form by the fact that the gas pressure distribution is primarily determined by the gravitationally dominant dark matter, in which the gas is in hydrostatic equilibrium, and whose density has been shown to approximately follow the NFW profile \citep*[e.g.][]{Navarro1997}. While gas tracing dark matter may be an acceptable approximation for relatively massive systems where the effects of feedback from AGN and supernovae are more minor, X-ray observations show the gas is significantly more extended (less cuspy) in low-mass groups (e.g.\ \citealt{Vikhlinin2006}). This behaviour is evident in our simulations with AGN feedback as well. At the very least, therefore, we anticipate that some of the parameters of the GNFW profile will need to be functions of halo mass, rather than fixed constants (as assumed in A10), in order to describe the pressure distribution of the hot gas over the full range of masses in the cosmo-OWLS AGN models. 

With the above in mind, we show, in Fig.~\ref{fig:Pmwfit}, the dimensionless mass-weighted pressure profiles for the \agn~8.0 model in eight different mass bins (increasing from top left to bottom right; the same bins were used for binning the tSZ flux in Fig.~\ref{fig:mmfcomparison}). We normalize the radii by $r_{500}$ and the pressures by $P_{500}$. Note that in order to further reduce the dynamic range of the $y$-axis, we plot $P/P_{500}(r/r_{500})^2$. Note also that the X-ray pressure profiles of the simulated groups and clusters from cosmo-OWLS have already been compared to the X-ray observations of \citet{Sun2011} (groups) and \citet{Arnaud2010} (clusters) by \citet{McCarthy2014}, who found that the fiducial AGN model (\agn~8.0) under consideration here reproduces the pressure profiles of both groups and clusters well. 

In Fig.~\ref{fig:Pmwfit}, the black filled circles with error bars correspond to the median profile in the corresponding mass bin and the error bars encompass 68 per cent of the systems. The solid blue, cyan, orange, magenta and red lines correspond to the universal pressure profile of the appendix B of \citetalias{Arnaud2010}, the best-fitting GNFW functional form with all five parameters left free to vary, the best-fitting GNFW form but with the concentration now a power-law of mass of the form $c_{500}=c_{500,0}(M_{500}/10^{14}~\textrm{M}_\odot)^\delta$, the best-fitting GNFW form but with the normalization now a power-law of mass of the form $P_{0}=P_{0,0}(M_{500}/10^{14}~\textrm{M}_\odot)^\epsilon$, and the best-fitting GNFW form but with the concentration and normalization now power-laws of mass, respectively. The bottom part of each panel shows the residuals of the best-fitting profiles. The functional forms are fitted simultaneously to the eight median profiles with their error bars (i.e.\ to the filled black circles with error bars). The best-fitting parameters for the final model which has both the concentration and normalization varying as power-laws of mass for the four cosmo-OWLS models used here and for both mass-weighted and median pressure profiles are listed in Table~\ref{table:Pfit}.  We note that some of the parameters are strongly degenerate with each other, but this is unimportant for our purposes since all that we require is a parametric model that reproduces simulated pressure profiles well.

\begin{figure}
\begin{center}
\includegraphics[width=1.0\hsize]{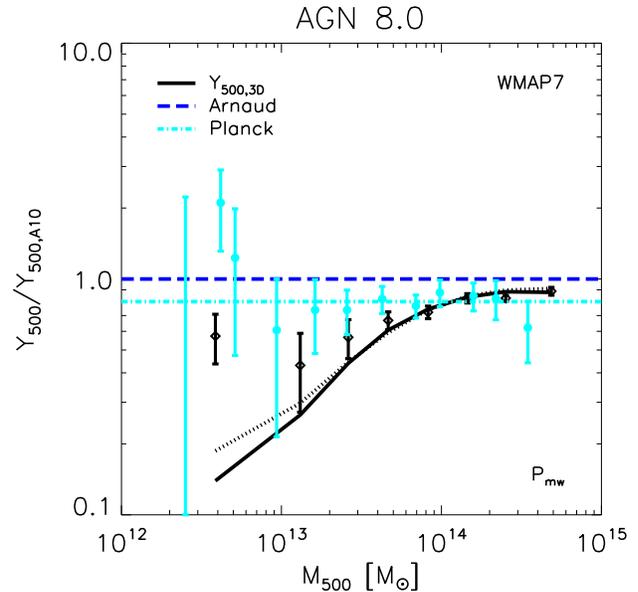}
\caption{Comparison of the $Y_{500}$--$M_{500}$ relations when a new mass-dependent spatial template based on the \agn~8.0 model is used in the matched filter. The blue dashed and cyan dot-dashed lines correspond to the best-fitting scaling relations of \citetalias{Arnaud2010} and \citetalias{Planck2013}, respectively. The solid and dotted black curves correspond to the unweighted and weighted mean {\it true} tSZ flux--$M_{500}$ relations for the \agn~8.0 model, respectively. The filled cyan circles and empty black diamonds with error bars correspond to the observational data of \citetalias{Planck2013} and the results of the synthetic observations (using the new template) of the simulated maps, respectively. The integrated tSZ fluxes have been normalized using the best-fitting scaling relation of \citetalias{Arnaud2010}. The recovered and true $Y_{500}$--$M_{500}$ approximately agree, implying the majority of the bias present in Fig.~\ref{fig:mmfcomparison} is due to the use of an inappropriate (fixed) spatial template.}
\label{fig:Pmwtest}
\end{center}
\end{figure}

From Fig.~\ref{fig:Pmwfit}, it is clear that the shape of the pressure profiles is indeed quite strongly mass-dependent (compare the black filled circles to the solid blue curves, which represent the fixed-shape universal pressure profile) and it is necessary to make two of the GNFW coefficients, the normalization and the concentration, mass-dependent in order to obtain a reasonable fit (solid red curve) over the whole radial and halo mass range when AGN feedback is included. Note that \citet{Battaglia2012} similarly found that the normalization and concentration as well as the outer slope $\beta$ needed to be not only mass-dependent but also redshift-dependent in order to provide a decent fit to the simulations of \citet{Battaglia2010} which include AGN feedback.  

\begin{figure*}
\begin{center}
\includegraphics[width=0.4898\hsize]{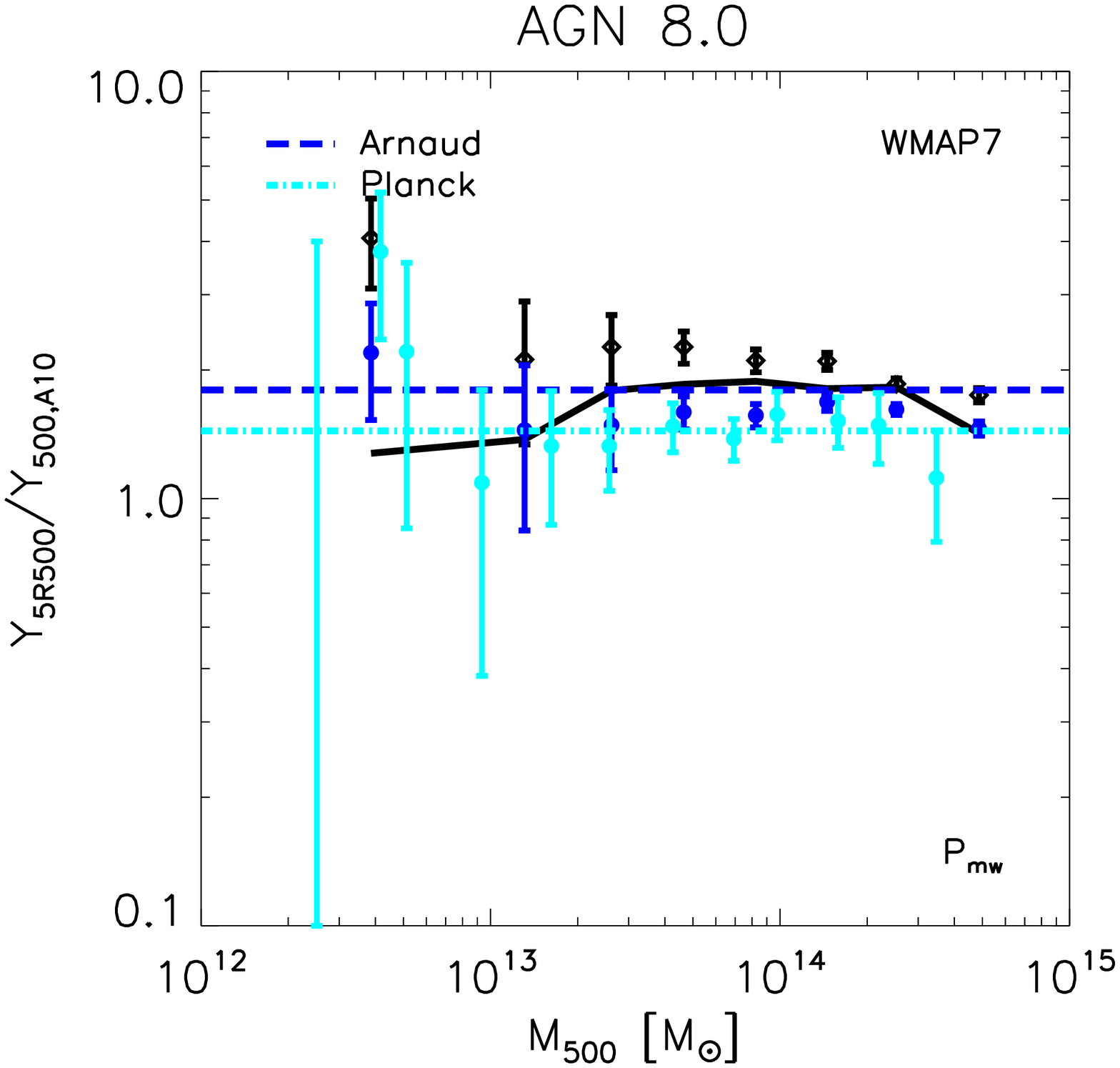}
\includegraphics[width=0.4898\hsize]{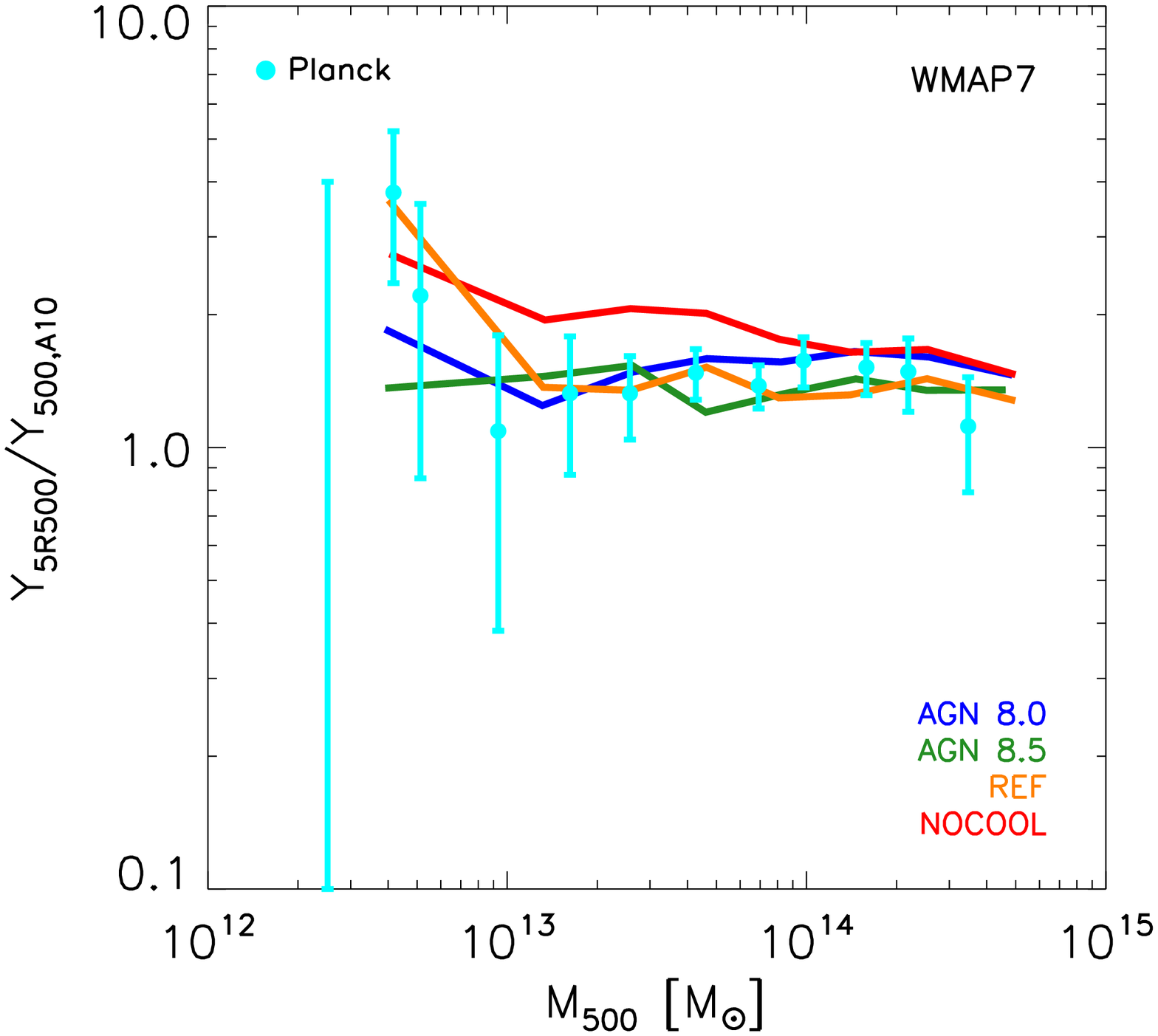}
\caption{Comparison of the $Y_{5r_{500}}$--$M_{500}$ relations. {\it Left:} The blue dashed and cyan dot-dashed lines correspond to the best-fitting scaling relations of \citetalias{Arnaud2010} and \citetalias{Planck2013}, respectively. The solid black curve correspond to the unweighted mean {\it true} $Y_{5r_{500}}$--$M_{500}$ relation for the \agn~8.0 simulation. The filled cyan circles, empty black diamonds and filled blue circles with error bars correspond to the observational data of \citetalias{Planck2013} and the results of the synthetic observations of the simulated maps with the extraction made using the new mass-dependent spatial template (derived from the \agn~8.0 run) and the standard template of \citetalias{Arnaud2010}, respectively. Overall, the level of bias present in the recovered flux within $5 r_{500}$ is relatively modest ($10-20$ per cent), independent of the choice of spatial template. {\it Right:}  A like-with-like comparison of the cosmo-OWLS runs with measurements of \citetalias{Planck2013}. The solid curves (red, orange, blue and green) represent the mean relations for the different physical models using the synthetic observations with the standard template. When comparing {\it observed} quantities (i.e. $Y_{5r_{500}}$ instead of $Y_{500}$), the \planck~observations actually exclude self-similar behaviour (\nocool) rather than support it. The fiducial AGN model (\agn~8.0), which reproduces the X-ray and optical properties of local groups and clusters, is consistent with the measured tSZ flux--mass trend.}
\label{fig:Pmwtest5r500}
\end{center}
\end{figure*}

We now proceed to alter the template adopted in the MMF, replacing the universal pressure profile with the template derived above (the solid red curve in Fig.~\ref{fig:Pmwfit}) and we repeat the recovery of tSZ fluxes from the synthetic maps. In Fig.~\ref{fig:Pmwtest}, we compare the results of the synthetic observations of the simulated maps for $z\le0.4$ based on the new template (empty black diamonds) with the mean spherically integrated (true) tSZ flux--$M_{500}$ relation (black curves), as well as the observational data (filled cyan circles with error bars). Note that when using the best-fitting pressure profile taken from the simulation as a spatial template in the matched filter, we numerically compute the conversion factor from $Y_{5r_{500}}$ into $Y_{500}$ using the new template profile for consistency.  In Appendix C, we show the derived conversion factor as a function of halo mass using the new template and compare it with the true conversion factor derived directly from the simulations.

Quite remarkably, the bias between the recovered and true values of $Y_{500}$ for the \agn~8.0 model in Fig.~\ref{fig:mmfcomparison} is largely removed. \emph{Hence, the majority of the bias between the recovered and true tSZ flux within $r_{500}$ is due to the use of an inappropriate spatial template in the matched filter.}  

Note that even with the new template, some residual bias remains, implying the template does not fully capture the detailed structure of the hot gas (e.g.\ halo-to-halo scatter may be significant, gas clumping, asphericity, etc.).  Indeed, as we show in Appendix C, the conversion factor (from $Y_{5r500}$ to $Y_{500}$) trend derived from the best-fitting mass-dependent spatial template does not perfectly match the true conversion factor relation from the simulations.  One may be able to derive a spatial template that better reproduces the true conversion factor from the simulations, but this would likely come at the expense of the template providing a poorer fit to the spherically-symmetric pressure profile.  Furthermore, even perfect knowledge of the conversion factor does not guarantee a bias-free estimate of $Y_{500}$, as the estimate of `total' flux ($Y_{5r500}$) may also be biased at some level, as we discuss immediately below.  

It is of interest to determine whether the removal of the bias (in Fig.~\ref{fig:Pmwtest}) with the use of the new template is due entirely to the improved conversion in the flux measured within $5 r_{500}$ to that within $r_{500}$, or if the `total' flux within $5 r_{500}$ itself has also been affected. That is, is the flux measured within the larger radius $5 r_{500}$ unbiased?  In the left panel of Fig.~\ref{fig:Pmwtest5r500}, we show the unweighted mean {\it true} $Y_{5r_{500}}$--$M_{500}$ relation for the \agn~8.0 simulation (solid lines), the observational data (filled cyan circles with error bars), the recovered relation from synthetic observations of the simulated maps for $z\le0.4$ when using the new halo mass-dependent spatial template (empty black diamonds), and the recovered relation when using universal pressure profile of \citetalias{Arnaud2010} (standard MMF; filled blue circles), respectively. Even when measured within $5 r_{500}$, the recovered fluxes are still biased at a relatively low level ($\sim10-20$ per cent) with respect to the true $Y_{5r_{500}}-M_{500}$ (solid line) for the AGN models. Using our new mass-dependent template improves the situation somewhat, but does not eliminate the bias altogether.  

For the sake of consistency and completeness, in the right panel of Fig.~\ref{fig:Pmwtest5r500}, we compare the mean $Y_{5r_{500}}-M_{500}$ relation for the synthetic observations using the standard MMF on the simulated maps for the different physical models (coloured solid curves) with the observational data of \citetalias{Planck2013} (filled cyan circles). This represents a like-with-like comparison between all the models and the observational data, in the sense that we are directly comparing what has actually been measured and the measurements for both the observations and the simulations have been made in a consistent manner (even if the standard MMF template has been shown to poorly reproduce the hot gas in some of these models). We conclude from this comparison that there is very good agreement between the trend measured by \citetalias{Planck2013} and that predicted by the fiducial AGN (\agn~8.0) model which reproduces the X-ray properties of local groups and clusters well (see \citetalias{LeBrun2014} and \citealt{McCarthy2014}). In contrast, the self-similar (\nocool) model \emph{is excluded} on the basis that it overpredicts the observed tSZ flux at low to intermediate halo masses.

\section{Summary and Discussion}
\label{sec:sum}

We have used synthetic observations produced from the cosmo-OWLS suite of cosmological hydrodynamical simulations to: (i) perform a consistent (like-with-like) comparison with recent observations; and (ii) to test the reliability of the observationally-inferred integrated tSZ flux--halo mass relation. While our methodology and tests have been geared towards a comparison with the trends recently reported in \citeauthor{Planck2013} (hereafter \citetalias{Planck2013}), the results and conclusions should be broadly applicable to tSZ flux measurements based on matched filter techniques.

From the analysis presented here, we reach the following conclusions:
\begin{enumerate}
\item Taken at face value, the results of PIntXI favour a close to self-similar $Y_{500}$--$M_{500}$ relation (Fig.~\ref{fig:Y500_M500}). Assuming that the gas has a temperature that is close to the halo virial temperature, such a trend would imply that the haloes of galaxies, groups and clusters all have a gas fraction within $r_{500}$ that is close to the universal baryon fraction and independent of mass. However, X-ray observations of local X-ray-bright galaxy groups (e.g.\ \citealt{Sun2009}) show that they are missing a large fraction of their baryons, indicating some inconsistency between the tSZ- and X-ray-derived results, as noted by \citetalias{Planck2013}.
\item We find that the multi-frequency matched filter (MMF) used by the \planck~collaboration recovers a nearly unbiased $Y_{500}$--$M_{500}$ relation for models which neglect efficient feedback (\nocool~and \refsim~; see top panels of Fig.~\ref{fig:mmfcomparison}). This is perhaps not too surprising (it is reassuring), since for these models the hot gas distribution has a shape similar to the universal pressure profile that is nearly independent of halo mass.
\item However, we find a significant bias in the recovered $Y_{500}$--$M_{500}$ relation (see bottom panels of Fig.~\ref{fig:mmfcomparison}) for models which invoke AGN feedback, which is necessary to reproduce the X-ray and optical properties of local groups and clusters (see \citealt{McCarthy2010,LeBrun2014}). The amplitude of the bias increases strongly with decreasing halo mass, reaching nearly an order of magnitude overestimate in $Y_{500}$ at halo masses below $\sim10^{13}$ M$_\odot$.
\item We have shown that the vast majority of the bias originates from the assumption of a fixed spatial template (the universal pressure profile), which becomes an increasingly poor description of the hot gas at low masses in our models with AGN feedback (Fig.~\ref{fig:Pmwfit}).
\item When a mass-dependent spatial template that describes the pressure profiles in the AGN models well is used (Fig.~\ref{fig:Pmwfit} and Table~\ref{table:Pfit}), the fluxes recovered for that simulation are nearly unbiased with respect to the true $Y_{500}-M_{500}$ relation of that simulation (Fig.~\ref{fig:Pmwtest}).
\item Encouragingly, fluxes measured within the larger aperture $5 r_{500}$ (i.e.\ $Y_{5r_{500}}$) only show modest levels of bias ($\sim10$ per cent), independent of the detailed shape of spatial template (left panel of Fig.~\ref{fig:Pmwtest5r500}) or the nature of the feedback implemented in the simulations (i.e.\ whether or not AGN feedback is included). Thus, the reported detection of a large fraction of the `missing baryons' in the form of hot halo gas by the \citetalias{Planck2013} appears to be quite robust, even if the gas is more extended than previously thought (see also \citealt{Greco2014}).
\item A consistent like-with-like comparison between the tSZ fluxes measured within the larger aperture $5 r_{500}$ demonstrates that the fiducial AGN model is in excellent agreement with the trends measured in \citetalias{Planck2013}, whereas the self-similar (\nocool) model actually overpredicts the tSZ flux at low to intermediate halo masses (right panel of Fig.~\ref{fig:Pmwtest5r500}).
\end{enumerate}

\begin{figure*}
\begin{center}
\includegraphics[width=0.33\hsize]{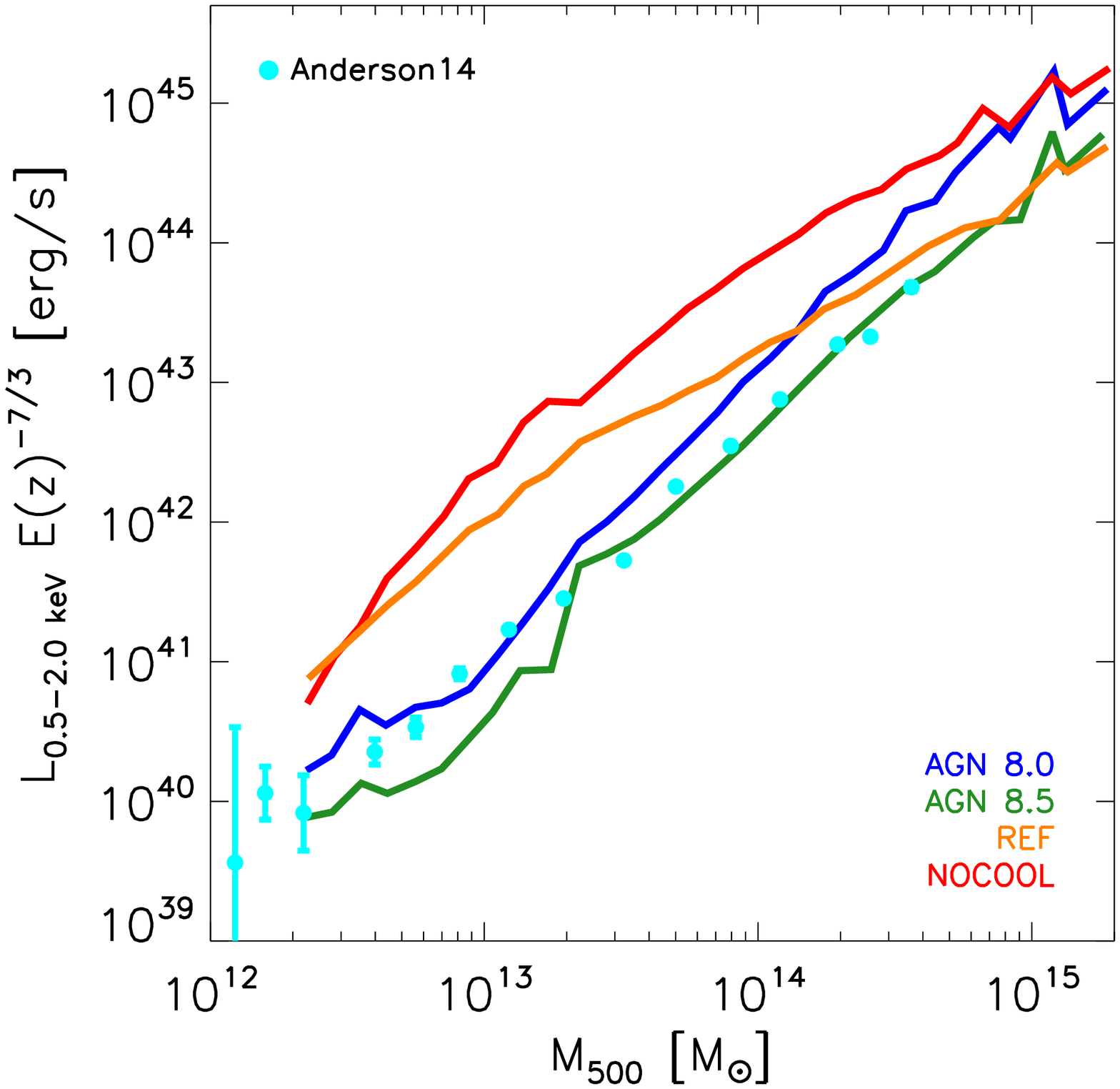}
\includegraphics[width=0.33\hsize]{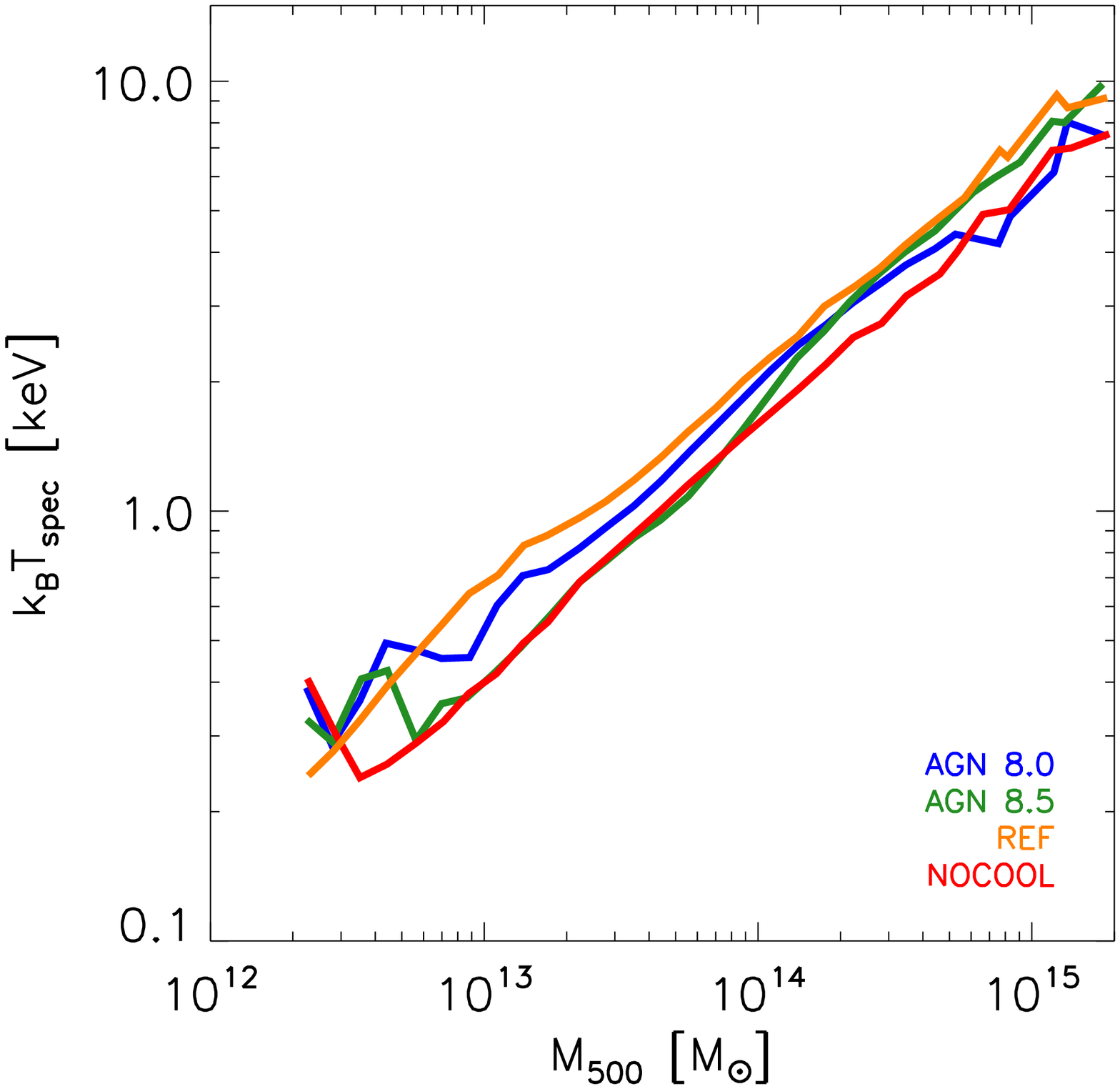}
\includegraphics[width=0.33\hsize]{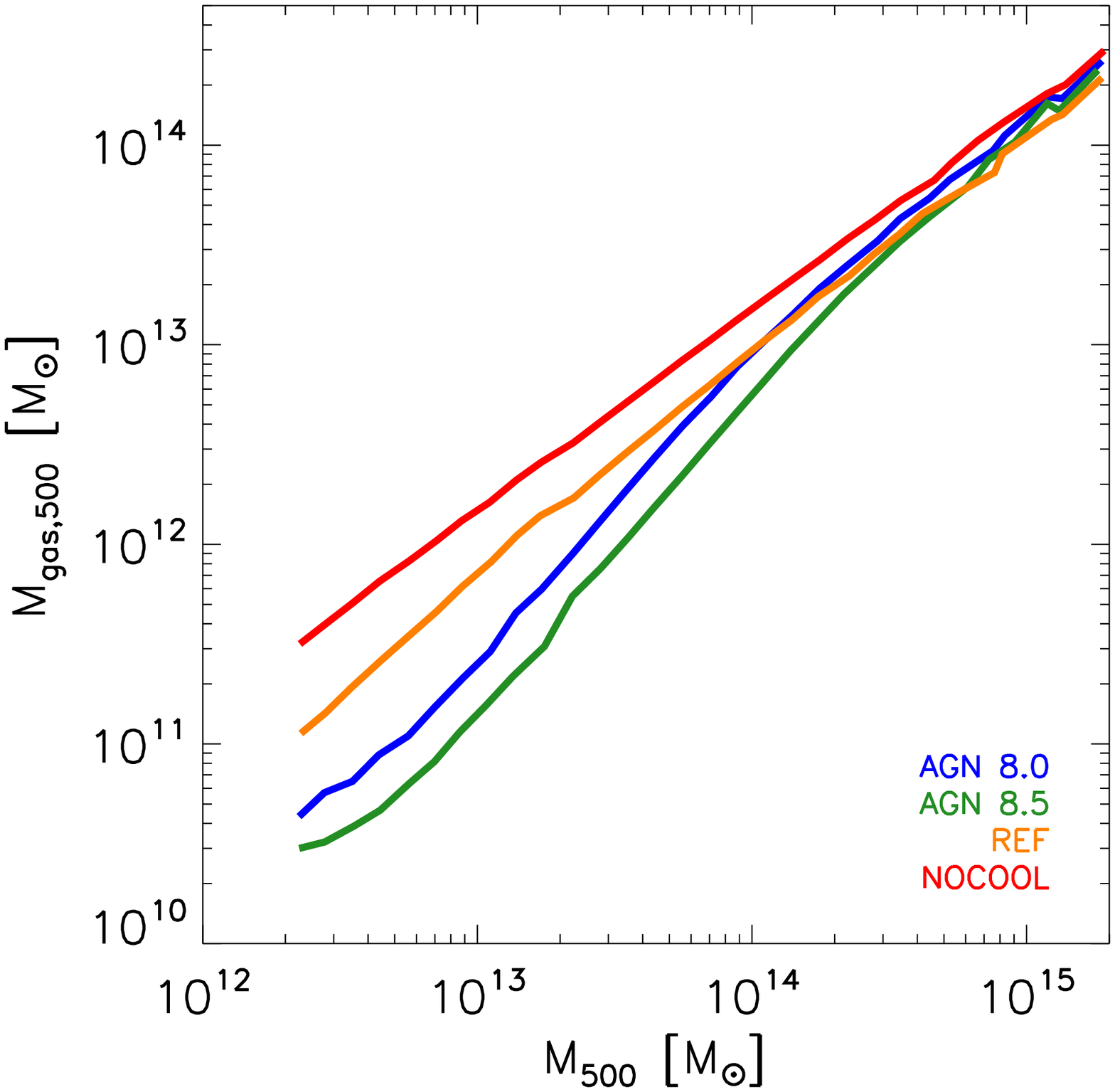}
\caption{The predicted $z=0$ soft X-ray (rest-frame $0.5-2.0$ keV) luminosity--, spectroscopic temperature--, and hot gas mass--halo mass mean relations for the different cosmo-OWLS runs.  The cyan points in the left panel represent the $L_{\rm X}$--$M_{500}$ relation of \citet{Anderson2015} inferred from a stacking analysis of RASS data of the SDSS LBG sample.  There is a clear steepening of the luminosity--halo mass and gas mass--halo mass relations due to gas ejection by AGN feedback.  The simulations with AGN feedback reproduce the observed $L_{\rm X}$--$M_{500}$ relation well.} 
\label{fig:Xray_M500}
\end{center}
\end{figure*}

The analysis of our AGN models indicate that the estimated tSZ fluxes within $r_{500}$ may be significantly biased (high) at low to intermediate masses.  However, it should be borne in mind that the estimate of the bias is quite sensitive to the nature of the feedback implemented in the simulations. We have used our current `best guess', as judged by previous comparisons of these models with resolved X-ray, tSZ, and optical observations (L14). Independent tests, however, are certainly well worthwhile. In particular, an obvious prediction of our results is that future high-resolution tSZ observations (e.g.\ SPTpol, ACTpol, SPT-3G, AdvACT, NIKA), which are capable of directly measuring the tSZ flux within $\theta_{500}$, ought to see a steepening in the $Y_{500}$--$M_{500}$ relation below $M_{500} \sim 10^{14}$ M$_\odot$ due to gas ejection by AGN.

An alternative test is to stack X-ray observations of a large optically-selected sample (such as the SDSS LBG sample in \citetalias{Planck2013}) and to directly measure the hot gas properties within $r_{500}$ in bins of mass for comparison to the tSZ-derived results.  \citet{Anderson2015} have recently undertaken such an exercise using the {\it ROSAT} All-Sky Survey (RASS).  They derive the stacked X-ray luminosity in bins of stellar mass and use the semi-analytic model of \citet{Guo2011} to infer the underlying $L_X$--$M_{500}$ relation.  Note that LBGs are well resolved with {\it ROSAT} (i.e.\ $\theta_{\rm PSF} \ll \theta_{500}$) and, in any case, the X-ray luminosity is centrally-concentrated due to the $\rho^2$ scaling of the X-ray emission.  \citet{Anderson2015} find a significantly steeper than self-similar relation and conclude that in order to reconcile their results with those of the \citetalias{Planck2013} study, the hot gas distribution must become increasingly extended (less concentrated) with decreasing halo mass.  This is qualitatively consistent with the behaviour predicted by our simulations with AGN feedback.

To be more quantitative, we plot in Fig.~\ref{fig:Xray_M500} the predicted mean soft X-ray (rest-frame $0.5-2.0$ keV) luminosity--, spectroscopic temperature--, and hot gas mass--halo mass relations at $z=0$ for the various cosmo-OWLS models.  Large differences are present in the predicted luminosities and hot gas masses at low to intermediate masses between the models which do and do not include AGN feedback.  

In the left-hand panel of Fig.~\ref{fig:Xray_M500}, we compare to the observed stacked X-ray luminosity--mass relation of \citet{Anderson2015}. For this comparison, we have scaled the observed relation to $z=0$ assuming self-similar evolution, using the mean redshift of each of the stellar mass bins. Consistent with the comparisons made in \citet{LeBrun2014}, we find that the observed $L_X$--$M_{500}$ related is well reproduced, both in terms of slope and zero point, by our AGN feedback models, with the \agn~8.0 and \agn~8.5 models effectively bracketing the data.

In any case, regardless of whether the simulations match the observations or not, {\it our findings highlight an important general conclusion: fluxes measured within large radii cannot reliably constrain the pressure profiles or integrated Y values within smaller radii when using matched filter-like techniques, because the obtained values depend directly on the assumed template profile}.  On the other hand, aperture photometry methods with aperture sizes approximately matched to the experiment under consideration should still work for these purposes (though likely at decreased signal-to-noise), so long as one does not try to extrapolate to smaller apertures.  {\it Planck} has proven to be exceptionally good at detecting large amounts of hot gas around massive galaxies (and has pushed much further down the halo mass scale than previous SZ experiments), thus confirming a fundamental prediction of cosmological structure formation.  The next step, however, is to map the distribution of the hot gas in detail and for this higher resolution experiments are required.

Finally, in terms of cosmological analyses with current {\it Planck} data, given that the flux measured within the larger radius $5 r_{500}$ can be reliably inferred independent of the details of the assumed spatial template, and that the simulations all predict very similar $Y_{5 r_{500}}$--$M_{500}$ relations nearly (although not completely) independent of the sub-grid details, it raises the question of whether the focus should shift to larger aperture measurements. If the goal is to lose sensitivity to the effects of uncertain baryonic physics in order to minimize systematic errors in cosmological analyses, this would indeed seem to be a sensible route to take in the future. One option would be to measure the tSZ flux within a large fixed physical aperture (to avoid covariances between the flux and mass) and to use simulations and/or a subset of the observations with very high quality data to calibrate the tSZ flux--mass relation needed for cosmological studies.

\section*{Acknowledgements}

The authors would like to thank the members of the OWLS team for their contributions to the development of the simulation code used here and the referee, Colin Hill, for a constructive report. AMCLB and IGM thank Simon White and Toby Marriage for helpful discussions. AMCLB is also grateful to Monique Arnaud and Gabriel Pratt for useful suggestions. AMCLB acknowledges support from an internally funded PhD studentship at the Astrophysics Research Institute of Liverpool John Moores University and from the French Agence Nationale de la Recherche under grant ANR-11-BD56-015. IGM is supported by an STFC Advanced Fellowship at Liverpool John Moores University.
This work used the DiRAC Data Centric system at Durham University, operated by the Institute for Computational Cosmology on behalf of the STFC DiRAC HPC Facility (www.dirac.ac.uk). This equipment was funded by BIS National E-infrastructure capital grant ST/K00042X/1, STFC capital grant ST/H008519/1, and STFC DiRAC Operations grant ST/K003267/1 and Durham University. DiRAC is part of the National E-Infrastructure. 

\bibliographystyle{mn2e}
\bibliography{stacking}

\appendix

\section{Resolution study}
\label{sec:res}

We examine the sensitivity of our results to numerical resolution. As currently available hardware prevents us from running higher resolution simulations in $400~h^{-1}$ Mpc on a side boxes, we use smaller simulations for testing numerical convergence and extending the mass range of the presented study (see Section~\ref{sec:owls}). They are $100~h^{-1}$ Mpc on a side and use $2\times256^3$ particles (which is the same resolution as our $2\times1024^{3}$ particles in $400~h^{-1}$ Mpc box runs) and $2\times512^3$ particles (i.e.\ eight times higher mass resolution and two times higher spatial resolution). They assume the \wmap7 cosmology. Note that the convergence tests are made using the true physical properties of the simulated systems (i.e.\ no synthetic observations were used).

\begin{figure}
\begin{center}
\includegraphics[width=1.0\hsize]{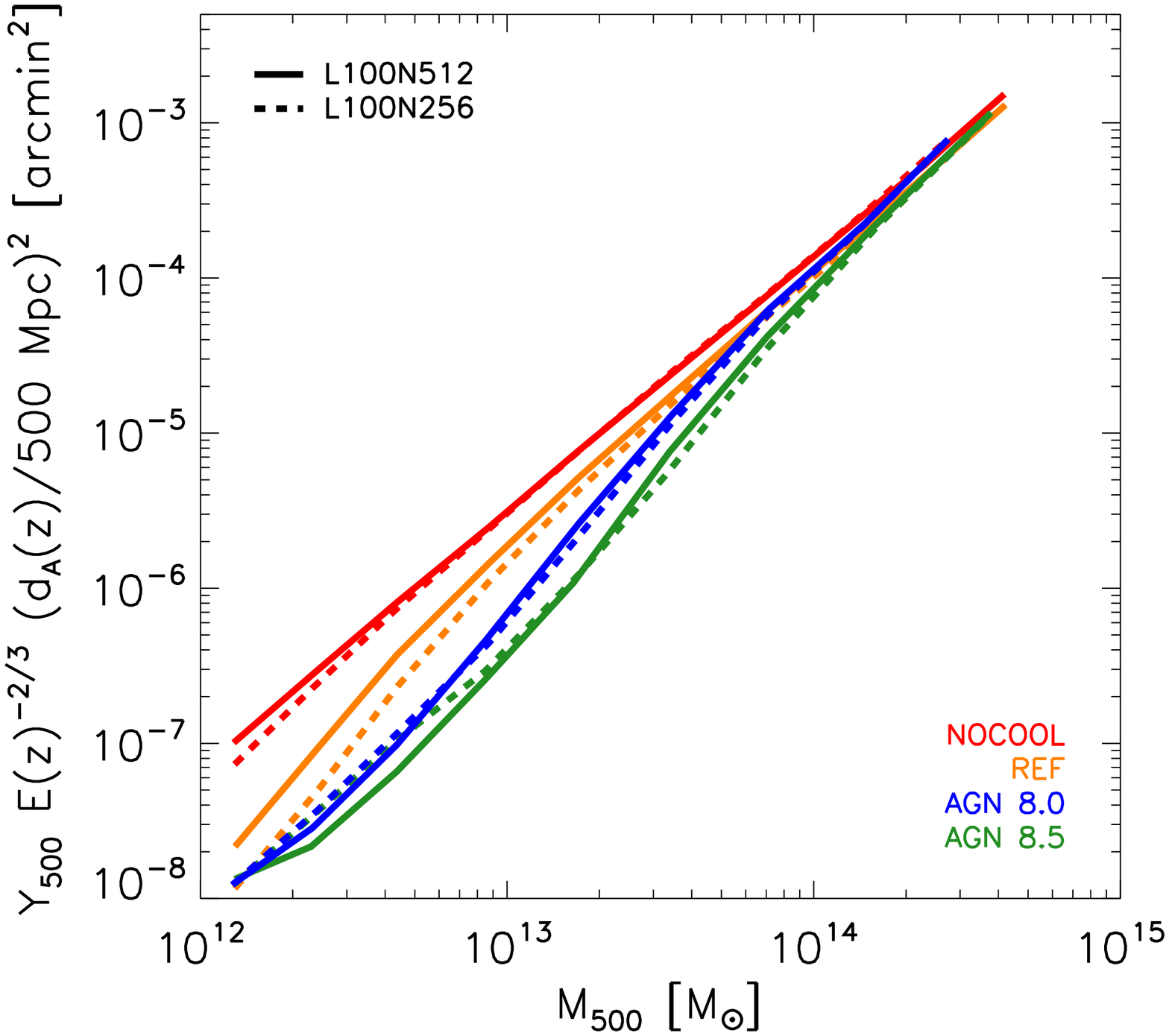}
\caption{Effect of numerical resolution on the median integrated tSZ flux--$M_{500}$ relation at $z=0$. The median integrated tSZ flux is adequately converged down to $M_{500}\sim2\times10^{13}~\textrm{M}_\odot$, thus justifying the transition between the high-resolution simulations and the production runs at $M_{500}=2\times10^{13}~\textrm{M}_\odot$.}
\label{fig:reso}
\end{center}
\end{figure}

In Fig.~\ref{fig:reso}, we compare the median integrated tSZ flux--$M_{500}$ relations at $z=0$ at the resolution of the production runs (dashed lines) and at eight times higher mass resolution (solid lines) for the four models used here (\nocool, \refsim, \agn~8.0 and \agn~8.5). We find that the integrated tSZ flux is adequately converged down to $M_{500}\sim2\times10^{13}~\textrm{M}_\odot$ at the resolution of the cosmo-OWLS production runs, justifying the transition between the high-resolution simulations and the production runs at $M_{500}=2\times10^{13}~\textrm{M}_\odot$ adopted in the present study. 

\section{Source Confusion}
\label{sec:confusion}

\begin{figure}
\begin{center}
\includegraphics[width=1.0\hsize]{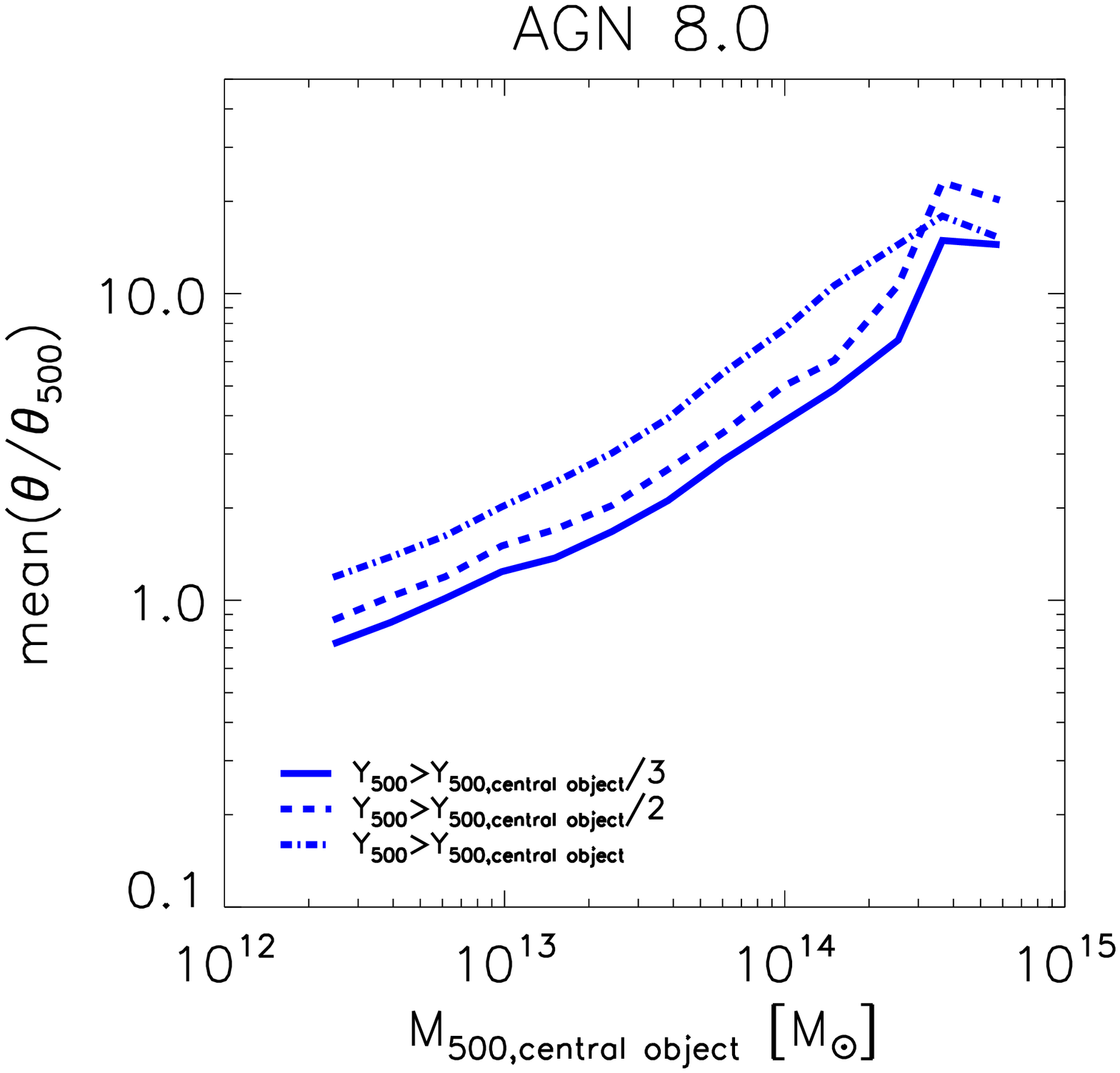}
\caption{The mean separation of bright background/foreground sources from the central halo (i.e.\ the object of interest) as a function of the central object's halo mass $M_{500}$. The angular separation is normalized by the angular size $\theta_{500}$ of the central object. The central objects are located at $z\le0.4$. The solid, dashed and dot-dashed curves correspond to the distance to objects with \smash{$Y_{500}>Y_{500,\textrm{central~object}}/3$}, \smash{$Y_{500}>Y_{500,\textrm{central~object}}/2$} and \smash{$Y_{500}>Y_{500,\textrm{central~object}}$}, respectively. If the fluxes are estimated within $5 r_{500}$, confusion is expected to become important for individual haloes with $M_{500} \la 10^{14}$ M$_\odot$.}
\label{fig:confusion}
\end{center}
\end{figure}

\begin{figure}
\begin{center}
\includegraphics[width=1.0\hsize]{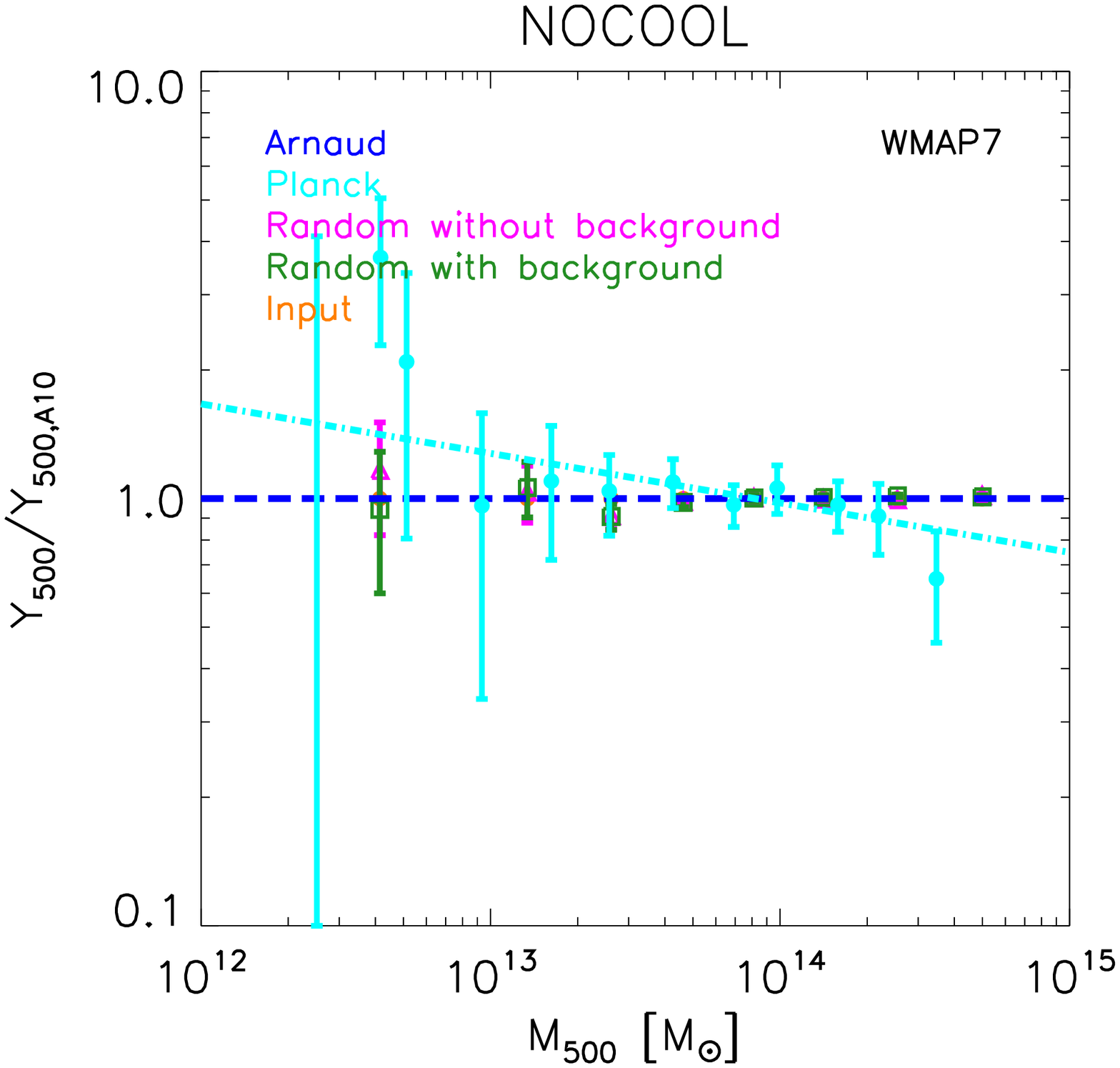}
\caption{Testing the effects of source confusion on the recovered mean $Y_{500}$--$M_{500}$ relation. The blue dashed and cyan dot-dashed lines correspond to the empirical best-fitting (see their equation 22) scaling relations of \citetalias{Arnaud2010} and \citetalias{Planck2013}, respectively. The filled cyan circles with error bars represent the observational data of \citetalias{Planck2013}. The filled orange circles correspond to the unweighted mean input $Y_{500}$ (computed using the empirical best-fitting $Y_{500}$--$M_{500}$ relation of \citetalias{Arnaud2010}). The empty green squares and magenta triangles with error bars correspond to the weighted mean and errors for the maps generated by injecting the GNFW haloes corresponding to the input $Y_{500}$ at random positions with and without using the original tSZ maps from the \nocool~simulation as background. The mean recovered flux is relatively insensitive to the presence of a tSZ background when a large number of haloes are stacked.}
\label{fig:confusiontest}
\end{center}
\end{figure}

The issue of source confusion in cluster surveys, and more specifically in tSZ surveys (for which source confusion is expected to be more problematic since the strength of the tSZ effect is redshift independent), especially those using instruments with rather large beams (the \planck~survey falls into that category with angular resolutions ranging from 5' to 31' depending upon the channel frequency) has already been explored using simulations of various degrees of realism over the past fifteen years (e.g.\ \citealt*{Voit2001,White2002}; \citealt{Hallman2007}; \citealt*{Holder2007}). In brief, they all concluded that confusion will be an issue for tSZ surveys of galaxies, groups and low-mass clusters ($M_{500}\lesssim10^{14}~\textrm{M}_\odot$). For instance, \citet{Voit2001} found using a back of the envelope calculation (using the Press--Schechter formalism) that the probability that any given line of sight will encounter a virialised structure with $k_BT\gtrsim0.5~\textrm{keV}$ is of order unity and that the virialised regions of groups and clusters cover over a third of the sky. Their use of the Hubble volume simulations \citep[e.g.][]{Jenkins2001}, which has the advantage over the analytic calculation of taking into account the clustering of virialised objects, corroborated their estimate of the group and cluster covering factor. 

In Fig.~\ref{fig:confusion}, we show the mean angular separation $\theta$ normalized by the angular size $\theta_{500}$ of the central object as a function of the central object's $M_{500}$. The central objects are located at $z\le0.4$ in order to mimic the LBG sample. The solid, dashed and dot-dashed curves correspond to the distance to objects with \smash{$Y_{500}>Y_{500,\textrm{central~object}}/3$}, \smash{$Y_{500}>Y_{500,\textrm{central~object}}/2$} and \smash{$Y_{500}>Y_{500,\textrm{central~object}}$}, respectively. The integrated tSZ flux $Y_{500}$ was computed approximately using the self-similar prediction: $Y_{500}\propto E(z)^{2/3}M_{500}^{5/3}/d_{A}^{2}$ where $d_{A}$ is the angular diameter distance. Both panels use the high-resolution version of the \agn~8.0 simulation below $M_{500}=2\times10^{13}~\textrm{M}_{\odot}$ and the production run above that threshold. Confusion is hence expected to be a source of large scatter (it could be more than a 100 per cent) at the low-mass end for estimates of the tSZ flux of individual haloes, as most of them have an object with a similar tSZ brightness (either in the foreground or the background) which overlaps with them in projection (i.e. $\theta/\theta_{500}<2$). Note that this can lead to both flux overestimation and underestimation as haloes with an overlapping neighbour with a comparable tSZ flux will have their flux boosted (could be up to doubled), whereas haloes with no overlapping neighbour with a comparable tSZ flux could have their flux underestimated (or even become undetectable) due to background overestimation.

While confusion is expected to result into sizeable errors in the recovered tSZ fluxes of individual low-mass haloes, the recovered mean Sunyaev--Zel'dovich flux (from stacking a large number of systems in mass bins) can still be unbiased. To test this hypothesis, haloes that were generated using a template based on the universal pressure profile of \citetalias{Arnaud2010} were injected into the original Compton $y$ maps. The results of this test are presented in Fig.~\ref{fig:confusiontest}. For each of the haloes in the synthetic LBG catalogue of the \nocool~simulation, the normalization of the profile (or equivalently $Y_{500}$) was set using the empirical $Y_{500}-M_{500}$ relation of \citetalias{Arnaud2010} (see their equation 22; dashed blue line). A new position was then drawn at random and the flux distributions from all the haloes in the catalogue were combined in order to generate a new Compton $y$ map on which the MMF was run both with and without using the original Compton $y$ maps of the \nocool~simulations as background. The filled orange circles correspond to the unweighted mean input $Y_{500}$ in bins of $M_{500}$. The empty green squares and magenta triangles with error bars correspond to the weighted mean and errors for the maps generated by injecting the generalized NFW (\citealt{Nagai2007}; see Section~\ref{sec:nonUPP}) haloes corresponding to the input $Y_{500}$ at random positions with and without using the original thermal tSZ maps from the \nocool~simulation as background. In order to limit the impact of the added noise (see Section~\ref{sec:mockLBG}), a 100 realisations of the small higher realisation simulation had to be used (instead of the usual 25) and 25 realisations (instead of the usual 2) were used for the large lower resolution simulation. The obtained results are consistent with there being no effects of uncorrelated confusion over the whole mass range (as the magenta and green symbols are consistent within their error bars with their input values) when a large number of haloes are stacked. The mean recovered flux is thus relatively insensitive to the presence of a tSZ background. This insensitivity to uncorrelated structures could be due to the fact that the MMF does not use the $k=0$ mode (which corresponds to the constant mean $y$ value of the map).

\section{$Y_{5r500}$ to $Y_{500}$ conversion factor}
\label{sec:conversion}

\begin{figure}
\begin{center}
\includegraphics[width=1.0\hsize]{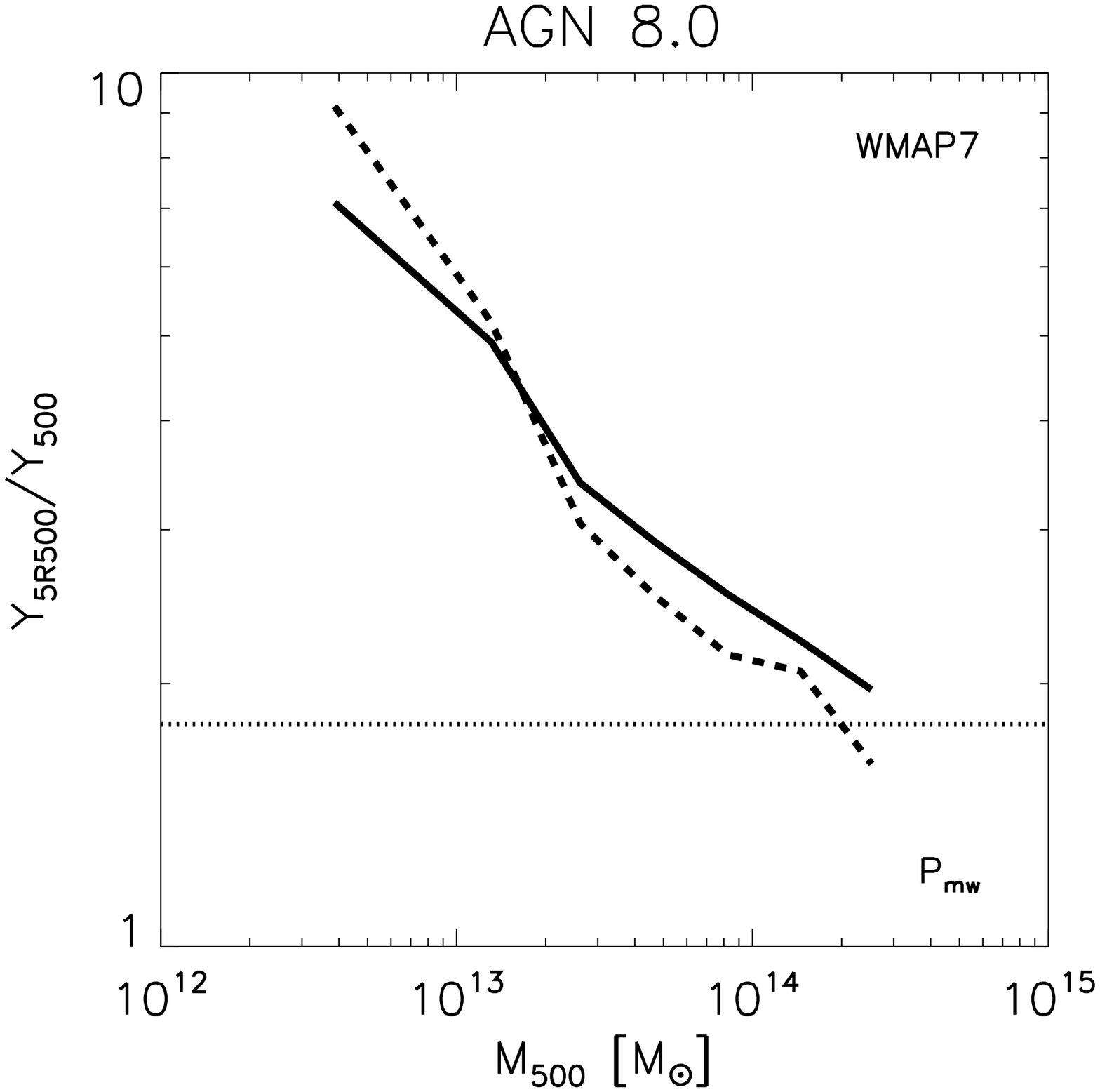}
\caption{Comparison of the conversion factor between $Y_{5r500}$ and $Y_{500}$ derived from the new mass-dependent spatial template (solid curve) with the true mean relation derived directly from the simulations (dashed curve).  The horizontal dotted line shows the conversion factor for the `universal pressure profile' of \citet{Arnaud2010}.  Overall the template reproduces the true trend well, but tends to slightly overestimate (underestimate) the conversion factor at the highest (lowest) masses.  The universal pressure profile significantly underestimates the magnitude of the conversion factor for galaxies and groups.}
\label{fig:conversiontest}
\end{center}
\end{figure}

In Fig.~\ref{fig:conversiontest}, we compare the conversion factor between $Y_{5r500}$ and $Y_{500}$ derived (numerically) from the new mass-dependent spatial template with the true mean relation derived directly from the simulations (i.e.\ using the full 3D particle distribution).  Overall the template reproduces the true trend well, but tends to slightly overestimate (underestimate) the conversion factor at the highest (lowest) masses.  Note that these small differences are apparent when one compares Fig.~\ref{fig:Pmwtest} and Fig.~\ref{fig:Pmwtest5r500} (left panel) in detail; compare the recovered fluxes (empty black diamonds) to the true relation (solid black curve) in the two plots and it is apparent that the inferred level of bias in the recovered flux differs slightly between the two plots.

\bsp
\label{lastpage}

\end{document}